\documentclass[twocolumn]{article}
\usepackage[utf8]{inputenc}
\usepackage[english]{babel}
\usepackage{subfig}
\usepackage{mwe}
\usepackage{xcolor}
\usepackage{graphicx}
\usepackage{amsmath}
\usepackage{physics}
\usepackage{bm}
\usepackage{stackengine}
\usepackage{indentfirst}
\usepackage{authblk}
\usepackage{url}

\title{Markov Chains for Modeling Complex Luminescence, Absorption, and Scattering in Nanophotonic Systems}
\author[1]{A. Ryan Kutayiah}
\author[1]{Smriti Kumar}
\author[1]{Rivi Ratnaweera}
\author[2]{Kenny Easwaran}
\author[1,3]{Matthew Sheldon}
\affil[1]{Department of Chemistry, Texas A\&M University, College Station, TX, USA}
\affil[2]{Department of Philosophy, Texas A\&M University, College Station, TX, USA}
\affil[3]{Department of Material Science and Engineering, Texas A\&M University, College Station, TX, USA }
\date{}                    
\graphicspath{ {./figures/} }

\begin{document}
\onecolumn
\maketitle

\begin{abstract}
    We develop a method based on Markov chains to model fluorescence, absorption, and scattering in nanophotonic systems. We show that the method reproduces Beer-Lambert's Law and Kirchhoff's Law, but also can be used to analyze deviations from these laws when some of their assumptions are violated.  We show how to use the method to analyze a luminescent solar concentrator (LSC) based on semiconductor nanocrystals.
\end{abstract}
\tableofcontents

\twocolumn

\section{Introduction}\label{introduction}
Recent progress in the study of nanophotonics, encompassing research topics such as metamaterials, plasmonics, and quantum confined matter has established that the optical interaction between light and matter is broadly tailorable across the electromagnetic spectrum via precise control of sub-wavelength scale optical structures \cite{Sharizal2018,Liu2011,Amendola2017,Yoffe2001,Baimuratov2012}.  Developments in radiation theory, aided by sophisticated computational methods, have informed experimental studies in nanophotonics, for example, as encapsulated in the inverse design approach \cite{Molesky2018} (and references therein).  Complex features of the optical response of nanoscale structures including scattering, absorption, emission, and nonlinear behavior are increasingly well understood \cite{Eisler2014,Sheldon2014,vandeGroep2016,LLi2016,Nadarajah2017,Pravitasari2018,Huang2018,Wu2018,Wu2019,Cheng2020,Martin2020}.

However, a significant outstanding challenge is the behavior of macroscopic systems composed of large assemblies of sub-wavelength nanophotonic elements.  In such systems complex interactions between individual elements on the nano- and mesoscale give rise to the macroscopic optical response. The challenge is compounded when these systems are optically pumped and driven far from equilibrium, so that simplifying assumptions commonly employed, such as Kirchhoff’s law for equating absorptivity and emissivity, are not \textit{a priori} guaranteed to be applicable \cite{Xiao2019,Xiao2020}.  Further, standard analytical or computational strategies, such as finite element methods, may be difficult or intractable to extend beyond nanoscale volumes or without simplifying boundary conditions \cite{Sullivan2000}. 

This work is focused on demonstrating how a computational strategy based on statistical analysis using Markov chains provides a powerful tool for modeling complex interactions between assemblies of nanophotonic elements that give rise to a macroscopic optical response. These interactions are mathematically described as pathways of radiative and non-radiative energy exchange within the optical system and between itself and the environment, including anisotropic absorption, photoluminescence, scattering, and loss. We show how this approach gives deep insight into the steady state behavior of the distribution of light and energy throughout the system, as well as other important optical properties, such as the angle-dependent emission or scattering from the assembly as a whole, even when the system is driven out of equilibrium. 

Markov chains have been used to model light-matter interactions before, particularly in the context of radiative transfer, for example, see \cite{Preisendorfer1965,Esposito1978}.  Mathematically, Markov chains also share some similarities with the more commonly used computational approach of Monte Carlo ray tracing.  While Monte Carlo ray tracing is popular, it is computationally taxing and it has been shown that the use of Markov chains applied to the same problems is computationally more efficient \cite{Li2016,Yang2018}. Monte Carlo methods approximate a distribution by sampling many instances from it, while the Markov Chain method allows us to calculate the distribution directly.

Moreover, we show how modeling nanophotonic assemblies using a Markov chain is also conceptually appealing.  The strategy is based on identifying the probabilities that connect the pathways of energy exchange between the nanoscale elements of the system, for example the angle-dependent emission or absorption profile of individual emitters. These descriptions of the optical behavior of individual elements in the ensemble can be obtained using other methods, for example Mie theory \cite{Schmid2014} or extended Mie theory \cite{Papoff2011}, finite element modeling \cite{Ussembayev2019} or finite-difference time-domain \cite{WSun2012}. Once the probabilities are identified the steady state behavior of the system is obtained simply by solving one eigenvalue problem.

This work is organized as follows: Section 2 describes the basics of a Markov chain for the kinds of systems we will consider. Section 3 fills in the details of these Markov chains to analyze the optical behavior of three specific systems.  The first two systems provide an analysis of Beer-Lambert's law and Kirchhoff's law, respectively.  The third models the behavior of a luminescent solar concentrator (LSC) comprised of semiconductor nanocrystals. Section 4 concludes the paper.

\section{Methods}\label{methods}
The primary method used to model optical interactions in this work is a probabilistic model known as a Markov chain.  First, we give a mathematical introduction to discrete-time Markov chains, Sec.~\ref{methods-markov chain}.  Then in Sec.~\ref{methods-model system}, we set up the specific Markov chain for the optical systems we analyze.  

\subsection{Markov Chains}\label{methods-markov chain}
 Using terms defined below, a discrete time, time-homogeneous Markov chain is an appropriate modeling device for a ``memoryless'' system in which the transitions at any moment depend only on the current state of the system, and not how the system has been at various points in the past, or how long it has been operating. Under certain simple conditions, the probability distribution of the system converges to a unique stationary distribution, that can be used to find the steady state of the system it represents. Our introduction to Markov chains loosely follows \cite{Klappenecker2018,Norris1997,Ross2007}.

A Markov chain is formally defined as a stochastic process that obeys the Markov property.
A \emph{stochastic process} is a collection of random variables $\mathcal{X}=\{X_t:t\in\mathcal{T}\}$ indexed by time. Here, we consider a discrete set of times $\mathcal{T}$. This ``time'' variable merely counts the number of elapsed transitions between states. We do not assume any correspondence of the number of time steps to a precise measurement of physical time in an experiment. The point of the model will not be to determine the precise time at which any event occurs, but rather the steady state distribution of the photons in the system. The development here assumes that the lifetime of each state is roughly equal, and that transition times are negligible, but the model can be modified to accommodate known divergences from these assumptions.

The random variables $X_t$ all take values in a finite set $\mathcal{S}$ of possible \emph{states}. Since the set of states is finite, the probability distribution for a random variable $X_t$ is determined by the probabilities of the individual states, $P(X_t=i)$. For convenience, we denote $P(X_t=i)$ by $d_i^{(t)}$, and the vector of probabilities for all states as $\vec{d}^{(t)}=\langle d_i^{(t)}\colon i\in\mathcal{S}\rangle$.

If the variables are all independent of each other, then the joint distribution $P((X_0=i_0)\&(X_1=i_1)\&(X_2=i_2)\&\dots)$ is just the product of the distributions of the individual variables $d_{i_0}^{(0)}d_{i_1}^{(1)}d_{i_2}^{(2)}\dots$. But a general stochastic process may have arbitrary dependencies among the variables, so that computing this joint distribution would require information about the conditional probabilities of each variable on every combination of the earlier ones. However, if the variables obey the \emph{Markov property}, then these conditional probabilities are determined by the value of the previous variable, and are independent of the values of all earlier variables. That is,
\begin{align}
    P(X_{t}=j|X_{t-1}=i,X_{t-2}=i_{t-2},\ldots,X_{0}=i_{0})&\nonumber\\
    = P(X_{t}=j|X_{t-1}=i).&\label{the Markov property}
\end{align}
Furthermore, if the process is \emph{time-homogeneous}, then this conditional probability is independent of $t$, and we can define \begin{equation}\label{time-homogeneous conditional probability}
    p_{ji}=P(X_{t}=j|X_{t-1}=i).
\end{equation}
This $p_{ji}$, the \emph{transition probability}, is the probability of the system \emph{transitioning} from state $i$ to state $j$.

By the Law of Total Probability, 
\begin{align}\label{a transition in probability form}
    P(X_{t}=j) & = \sum_{i\in\mathcal{S}}P((X_{t}=j)\&(X_{t-1}=i)) \\
    & = \sum_{i\in \mathcal{S}}P(X_{t}=j|X_{t-1}=i)P(X_{t-1}=i).
\end{align}
or 
\begin{equation}\label{a transition in matrix element form}
    d_{j}^{(t)} = \sum_{i\in\mathcal{S}}p_{ji}d_{i}^{(t-1)}.
\end{equation}
If we think of $\vec{d}^{(t)}$ as a column vector and let $\bm{P}$ be the matrix with $p_{ji}$ in the $j$th row and $i$th column, called the \emph{transition matrix}, then this means
\begin{equation}\label{concise iterative markov chain}
    \vec{d}^{(t)}=\bm{P}\vec{d}^{(t-1)}. 
\end{equation}
Since the columns of $\bm{P}$ sum to 1 (as they represent the probabilities of all values of $X_t$ conditional on a particular value of $X_{t-1}$), it is said to be a \emph{stochastic matrix}. %(In Refs.~\cite{Klappenecker2018,Norris1997,Ross2007}, they consider row vectors with matrix multiplication on the right, and a stochastic matrix is one in which the rows sum to 1.)

A Markov chain is said to be \emph{reducible} if there are states $i$ and $j$ such that, if one of the variables ever takes value $i$, then no later variable ever takes the value $j$, and \emph{irreducible} otherwise. It is said to be \emph{periodic} if there is some integer $n>1$ such that $X_t=X_{t'}$ only if $|t-t'|$ is a multiple of $n$, and \emph{aperiodic} otherwise. All our Markov chains are irreducible and aperiodic.

Each vector $\vec{d}^{(t)}$ represents the probability distribution of the system at a time. If there is a distribution $\vec{d}^{(s)}$ with
\begin{equation}\label{eigenvalue problem with gamma equal one}
    \bm{P}\vec{d}^{(s)}=\vec{d}^{(s)},
\end{equation}
then it is said to be a \emph{stationary distribution} of the system. For an irreducible, aperiodic Markov chain, there is always a unique stationary distribution \cite{Klappenecker2018,Norris1997,Ross2007}, and furthermore, for any initial distribution $\vec{d}^{(0)}$, we have
\begin{equation}\label{long-run convergence of markov chain}
    \lim_{t\rightarrow\infty}\bm{P}^{t}\vec{d}^{(0)} = \vec{d}^{(s)}.
\end{equation}
Thus, for such a Markov chain, the stationary distribution represents the steady state behavior of the system, and we can find the stationary distribution by solving Eq.~(\ref{eigenvalue problem with gamma equal one}). Note that this  is an eigenvalue equation with an eigenvalue of 1.

\subsection{Model}\label{methods-model system}

We first introduce a system with some simplifying assumptions in order to establish a concrete connection between the Markov chain formalism (Sec.~\ref{methods-markov chain}) and fluorescence, scattering, and absorption. In this work, a \textit{system} comprises a medium and an environment, which interact with each other via photon exchange. The environment sends photons into the medium at a constant rate and angular profile.    

The medium could be a solid such as a semiconducting slab (Sec.~\ref{results-beer-lambert law}), a colloidal suspension of nanoparticles (Sec.~\ref{results-beer-lambert law}), or nanoparticles embedded in a solid polymer matrix (Sec.~\ref{results kirchhoffs law} and \ref{results lsc}). In this section, the medium is a solid matrix of uniform refractive index $n_m$ that hosts uniformly distributed fluorophores. The environment around the medium has refractive index $n_{e}$. 

The emission of the fluorophores is assumed to be azimuthally symmetric. Thus, we represent angle of emission just by the zenith angle $\theta$, and assume emission into any angle is proportional to some function $f(\theta)$.

We consider two kinds of fluorophores and label them as isotropic or dipolar.  By an isotropic fluorophore, we mean that the fluorophore emits radiation with equal probability in all directions; we also assume that there is no anisotropy in its absorption cross-section $\sigma$.  Thus, for isotropic fluorophores, $f(\theta)=1$.  In contrast, the dipolar fluorophores display anisotropy in the emission and absorption-cross section, specifically $f(\theta)=\sin^{2}\theta$, see Ref.~\cite{Baimuratov2012}; and $\sigma(\theta)\propto\sin^{2}\theta$, see Ref.~\cite{Hens2012}. Further, all of the dipolar fluorophores emit radiation with the dipole oriented along the same axis, normal to the top surface of the medium. We will later draw a parallel between this behavior and ensembles of homeotropically aligned nanorod-shaped semiconductor nanocrystals in LSC-like devices. 

We also assume that the light emitted by either type of fluorophores is monochromatic. At first we assume light coming from the environment is the same wavelength as the fluorophore emission, but in Sec.~\ref{results lsc} this assumption is relaxed. The host material of the medium is non-absorbing, and the real part of the fluorophores' refractive index is matched to that of the host medium. We assume the fluorophores are far enough apart so that  near-field effects and resonant energy transfer can be neglected in the probabilistic description for energy transfer. Finally, we assume that the medium is of infinite extent in the $x$- and $y$- directions and has a finite thickness $L$ in the $z$-direction.

Now we are in a position to describe the state space $\mathcal{S}$, random variables $X_t$ with distributions $\vec{d}^{(t)}$, and transition matrix $\mathbf{P}$, as they relate to this system. The Markov chain tracks the movement of an individual photon through the system. For a system with multiple photons, the stationary distribution of each photon will be the same, and the observed distribution of photons in the steady state will be proportional to this distribution.

The state space $\mathcal{S}$, is constructed by stratifying the medium into $M$ equal thickness parallel slabs.  We call these slabs \emph{layers}.  For a medium of thickness $L$, a layer has thickness $\Delta z=L/M$. Each layer is treated as a state representing a possible location for the photon at a time. In all our simulations reported here, we use $M=1000$. We found that increasing $M$ to $10,000$ only made a difference to the results on the order of $10^{-8}$.

$X_t$ represents the ``position" of a photon at a time $t$, where positions are layers.  Note that we only track the position of a photon along the $z$-axis since $X_{t}$ takes values from $\mathcal{S}$ which is created by discretizing the system along $z$.  It does not account for the location of the photon in the $x$ and $y$ directions. Our standard assumption is that the systems are large enough and homogeneous enough in these horizontal dimensions that we can treat them as infinite, though see further discussion in Sec.~\ref{results lsc}.

The probability $d^{(t)}_{i}=P(X_{t}=i)$, represents the probability that a photon had its most recent absorption event at layer $i$ at time $t$, that is, the probability that a photon is ``located at'' this layer. In describing a system with multiple photons, we will sometimes refer to $d^{(t)}_{i}$ as the population (or number of photons) of layer $i$.   

A transition probability $p_{ji}$ of the system is identified as the probability that a photon from layer $i$ at time $t-1$ moves to layer $j$ at time $t$.  To move from layer $i$ to layer $j$, a photon must be emitted from layer $i$, transmitted through all intervening layers, and absorbed at layer $j$.  In this simplified model the probabilities of these three events depend only on the zenith angle at which the photon is emitted; thus, we calculate each of these three probabilities at each angle, multiply them, and integrate over all possible angles of travel between these two layers. (As discussed in Sec.~\ref{methods-model system-transmission probability}, these are the angles from 0 to $\pi/2$.) Denoting the emission probability density as $dp^{(e)}(\theta)$, the transmission probability as $p^{(t)}(\theta,i,j)$ and the absorption probability as $p^{(a)}(\theta)$, the transition probability $p_{ji}$ can be written as
\begin{equation}
    p_{ji} = \int_{0}^{\pi/2} dp^{(e)}(\theta)p^{(t)}(\theta,i,j)p^{(a)}(\theta).\label{general transition probability matrix element with azimuthal symmetry}
\end{equation}
We describe $dp^{(e)}(\theta)$ in greater detail in Sec.~\ref{methods-model system-emission probability}, $p^{(a)}(\theta)$ in Sec.~\ref{methods-model system-absorption probability}, and $p^{(t)}(\theta,i,j)$ in Sec.~\ref{methods-model system-transmission probability}. In Sec.~\ref{methods-model self-absorption} we discuss the special features of the case where $i=j$.
%Eq.~(\ref{general transition probability matrix element with azimuthal symmetry}) holds for $i\neq j$. We will consider the special case when $i=j$ after we specify the probability functions $dp^{(e)}(\theta)\text{, }p^{(t)}(\theta,i,j)\text{, and }p^{(a)}(\theta)$ in more detail.
The transition probabilities and states are pictorially represented by a transition diagram in Fig.~\ref{schematic of processes}. Importantly, since the emission probabilities are assumed to be proportional to $f(\theta)$, and the transmission probability depends only on the path traversed from one layer to the next, these transition probabilities do not depend on any aspect of the system other than what layer the photon was most recently absorbed by, so the system obeys the Markov condition. 

In the physical system we are modeling, photons are sent into the system at a constant rate, move according to the transition probabilities from layer to layer within the medium, and eventually escape. While it does have a unique steady state, such a system is not strictly a Markov process, because of the entry and escape of photons from the set of states. However, for any such system, there is a related fictitious system with the same transition probabilities among the states in the medium, but with an extra dummy state.  Photons entering the medium transition from the dummy state to the medium, while photons escaping the medium transition from the medium to this dummy state. That is, in this fictitious system the photons entering the medium \emph{are} the photons that just escaped. This fictitious system is a Markov process because the photon number is conserved, and therefore has a unique, identifiable steady state distribution. 

This fictitious system does not behave like the physical system outside the steady state; in the fictitious system the number of photons entering the system at a time $t$ is the number that escaped at time $t-1$, while in the physical system it is constant. However, the steady state distributions must be the same. Both systems have the same transition probabilities among the states within the medium. Furthermore, in the steady state of the physical system (assuming for now no non-radiative loss, but see Sec.~\ref{methods-model system-nonunity quantum yield}), the number of escaping photons will equal the number of entering photons, just like in the fictitious system. Thus, our analysis focuses on finding the steady state of this fictitious Markov process, which is also the steady state of the physical system.

We refer to the dummy state as ``the environment''. The entire set of states is $\mathcal{S}=\{0,1,\ldots,M\}$, where layer $0$ is the environment, while the remaining layers, $1,\ldots,M$, belong to the medium, see Fig.~\ref{fig: general diagram and markov graph and transmission distance}. Transition probabilities from and to the ``environment'' layer are discussed in Secs.~\ref{methods-model environment to medium} and \ref{methods-model to environment} respectively.

The steady state distributions $\vec{d}^{(s)}$ are eigenvectors, and thus can be scaled by any constant. We will always display steady state distributions scaled so that the population of the ``environment'' layer is 1. This can be interpreted physically as treating all modeled systems as having the same intensity of light incident on them, and comparing the relative populations of the layers of the different systems under this condition. The populations of the layers are sensitive to how thinly the medium is sliced into these discrete layers, but the trend will be the same.  Thus, the normalized steady state distribution is given by

\begin{equation}\label{steady-state vector normalized by environment}
    \vec{d}^{(s)} \rightarrow \left\langle 1,\frac{d^{(s)}_{1}}{d^{(s)}_{0}},\ldots, \frac{d^{(s)}_{M}}{d^{(s)}_{0}}\right\rangle.
\end{equation}

\subsubsection{Emission Probability Density}\label{methods-model system-emission probability}
 
We develop the emission probability density by first considering emission from a layer inside the medium.  We will discuss emission from the environment to the medium on a case-by-case basis as different scenarios are considered in Secs.~\ref{results-beer-lambert law}, \ref{results kirchhoffs law}, and \ref{results lsc}.  

For a layer of the medium, the emission probability density $dp^{(e)}(\theta)$ is defined as the probability that a photon is emitted at an angle whose zenith is infinitesimally close to $\theta$. If $f(\theta)$ is proportional to the probability of emission in solid angle $d\Omega=\sin\theta d\theta d\phi$, then integrating over azimuth angle $\phi$ yields 
\begin{equation}\label{general emission probability density}
    dp^{(e)}(\theta)=2\pi\nu f(\theta)\sin\theta d\theta,
\end{equation}
where $\nu$ is a normalization constant to ensure that the total probability is 1. That is,
\begin{equation}\label{general normalization for medium}
    \nu = \frac{1}{\int_0^\pi 2\pi f(\theta)\sin\theta d\theta}.
\end{equation}
For isotropic fluorophores, $f(\theta)=1$ and $\nu=1/4\pi$. For dipolar fluorophores, $f(\theta)=\sin^2\theta$ and $\nu=3/8\pi$.

\subsubsection{Absorption Probability}\label{methods-model system-absorption probability}

For a photon traveling at angle $\theta$ that starts outside a layer and then reaches it, the probability of being absorbed by that layer can be determined from the well-known Beer-Lambert law \cite{Swinehart1962} (see Ref.~\cite{Waldenfels2011} for a derivation of Beer-Lambert's law based on stochastics). Recall that $\Delta z$ is the thickness of the layer, so the path length through the layer along this angle is $\Delta z/|\cos\theta|$. Let $\alpha(\theta)=\rho\sigma(\theta)$, where $\rho$ is the fluorophore concentration and $\sigma(\theta)$ is the absorption cross-section (which has angle dependence in the case of anisotropic fluorophores). Then, the probability of transmitting through the layer is $e^{-\alpha(\theta)\Delta z/|\cos\theta|}$ while the probability of being absorbed by the layer is given by
\begin{equation}\label{general absorption probability}
    p^{(a)}(\theta) = 1-e^{-\alpha(\theta)\Delta z/|\cos\theta|}.
\end{equation}

\subsubsection{Transmission Probability}\label{methods-model system-transmission probability}

For a photon that has been emitted by one layer $i$ at an angle $\theta$, to find the probability of it being transmitted to another layer $j$, we calculate the probability of it \emph{not} being absorbed along the path from layer $i$ to layer $j$. The length of the full path the photon travels before reaching layer $j$ depends on where precisely in layer $i$ the photon was emitted. We approximate by assuming that all emissions occur from the center of the layer, and divide the medium into a large number of layers to ensure that this is a good approximation.

The path from layer $i$ to layer $j$ can be direct, or can involve reflections. In all our models, the top surface of layer 1 is an interface between the medium and the environment, and reflection occurs if $\theta$ is outside the escape cone; the escape cone is formed by the critical angle $\theta_{c}$ which is determined from Snell's law. In Sec.~\ref{results-beer-lambert law}, the bottom surface of layer $M$ is also an interface between the medium and the environment that reflects at these angles, while in Sec.~\ref{results kirchhoffs law} and \ref{results lsc} the bottom surface is an ideal mirror that reflects at any angle.

These reflections change the angle of travel from $\theta$ to $\pi-\theta$ and vice versa. But in all our systems, $dp^{(e)}(\theta)=dp^{(e)}(\pi-\theta)$ and $p^{(a)}(\theta)=p^{(a)}(\pi-\theta)$. Thus, in these calculations, we will always treat $\theta$ as a number between $0$ and $\pi/2$, understanding that it might represent the angle from the vertical in either direction, so that reflections do not change it.

\begin{figure}[!ht]
\centering
    \begin{tabular}[t]{r}
    %\hline
     \subfloat[\label{multiple reflections illustration}]{%
       \includegraphics[width=0.22\textwidth]{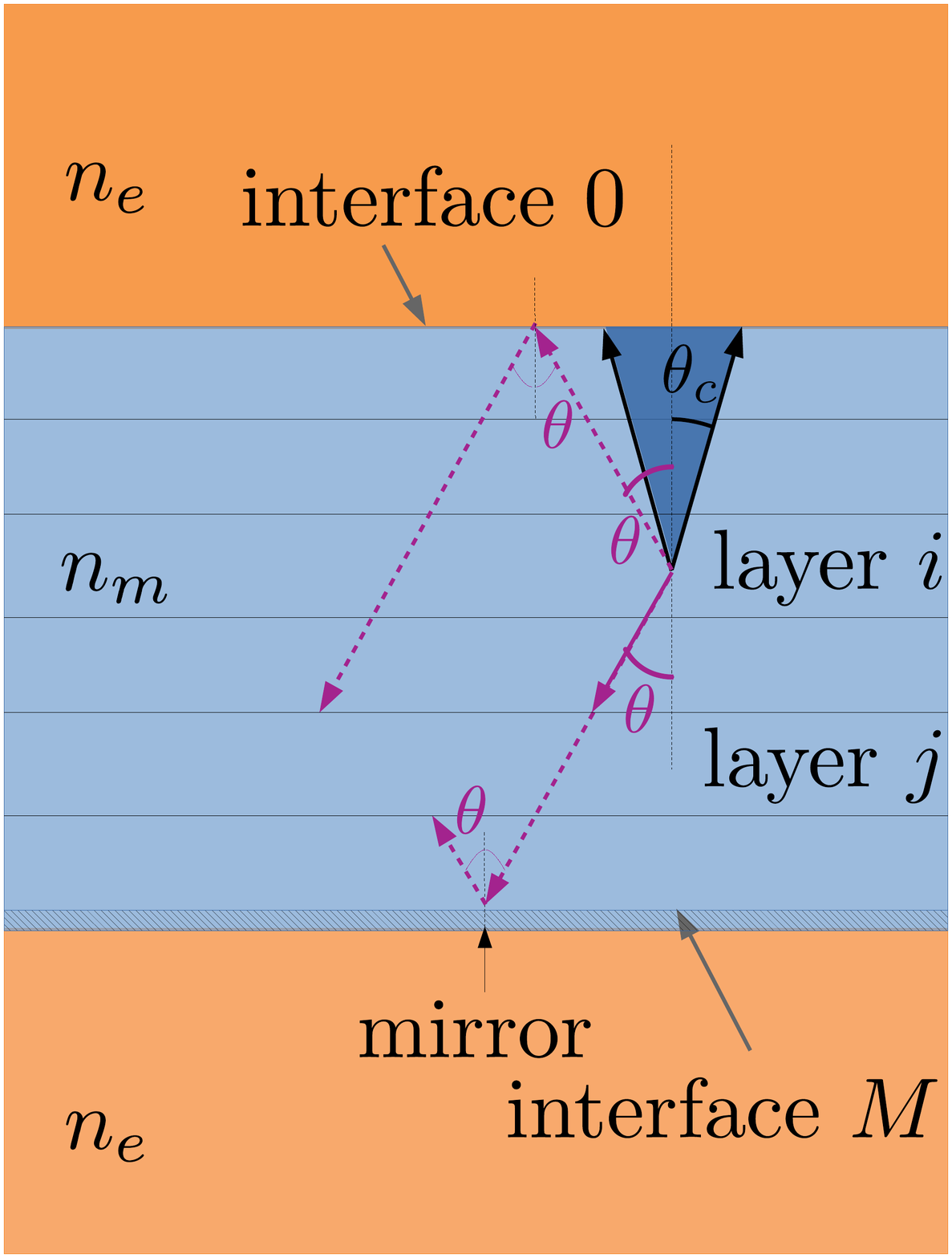}
     }
%     \\
%\hline
     \subfloat[\label{multiple reflections probability tree}]{%
      \includegraphics[width=0.22\textwidth]{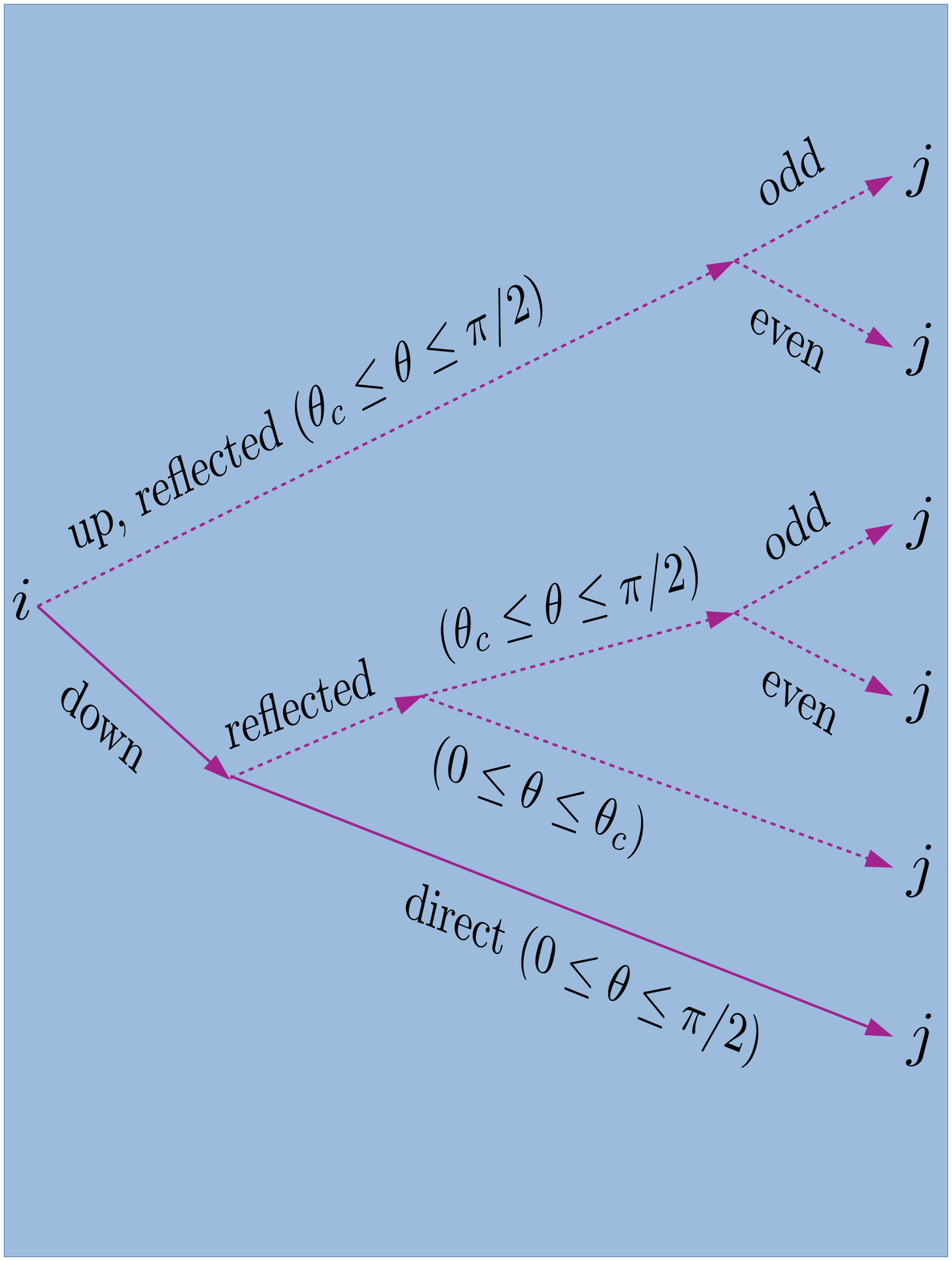}
     }\\
     \end{tabular}

     \caption{(a)  Illustration showing transitions from layer $i$ to $j$ by a direct path (solid arrow) and by reflections (dashed arrows).  (b)  A probability tree representing the different paths a photon, first emitted into the lower hemisphere, can take depending on the angle at which it is emitted in that hemisphere.}
     \label{fig:probability tree}
\end{figure}

Because direct paths and reflected paths are two separate ways for a photon emitted at angle $\theta$ from layer $i$ to reach layer $j$, the transmission probability is the sum of the direct and reflection transmission probabilities
\begin{equation}\label{total transmission probability}
    p^{(t)}(\theta,i,j)=p^{(t)}_{D}(\theta,i,j)+p^{(t)}_{R}(\theta,i,j).
\end{equation}
Figure~\ref{multiple reflections illustration} depicts the paths associated with the direct and reflection (one reflection) transmission probabilities.  Figure~\ref{multiple reflections probability tree} shows a probability tree for the transmission probability. 
The total transmission probability Eq.~(\ref{total transmission probability}) is what goes into the computing the transition probability from layer $i$ to $j$ using Eq.~(\ref{general transition probability matrix element with azimuthal symmetry}). However, as we observe at the end of this section, much of the reflection term is negligible in cases where the medium is thick and highly absorbing.

\begin{figure}[!ht]
\centering
    \begin{tabular}[t]{r}
    %\hline
     \subfloat[\label{Markov graph general}]{%
       \includegraphics[width=0.22\textwidth]{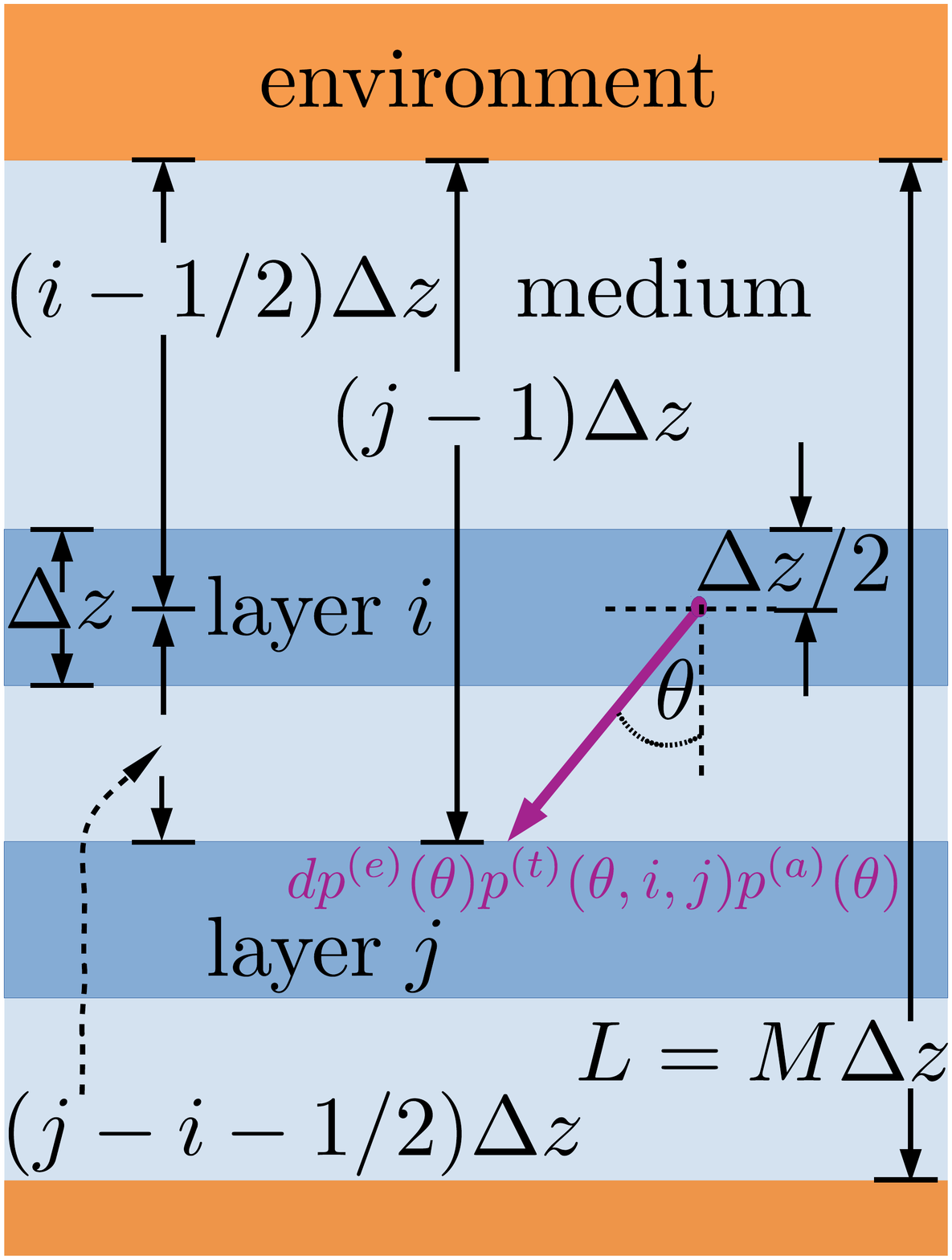}
     }
%     \\
%\hline
     \subfloat[\label{schematic of processes}]{%
      \includegraphics[width=0.22\textwidth]{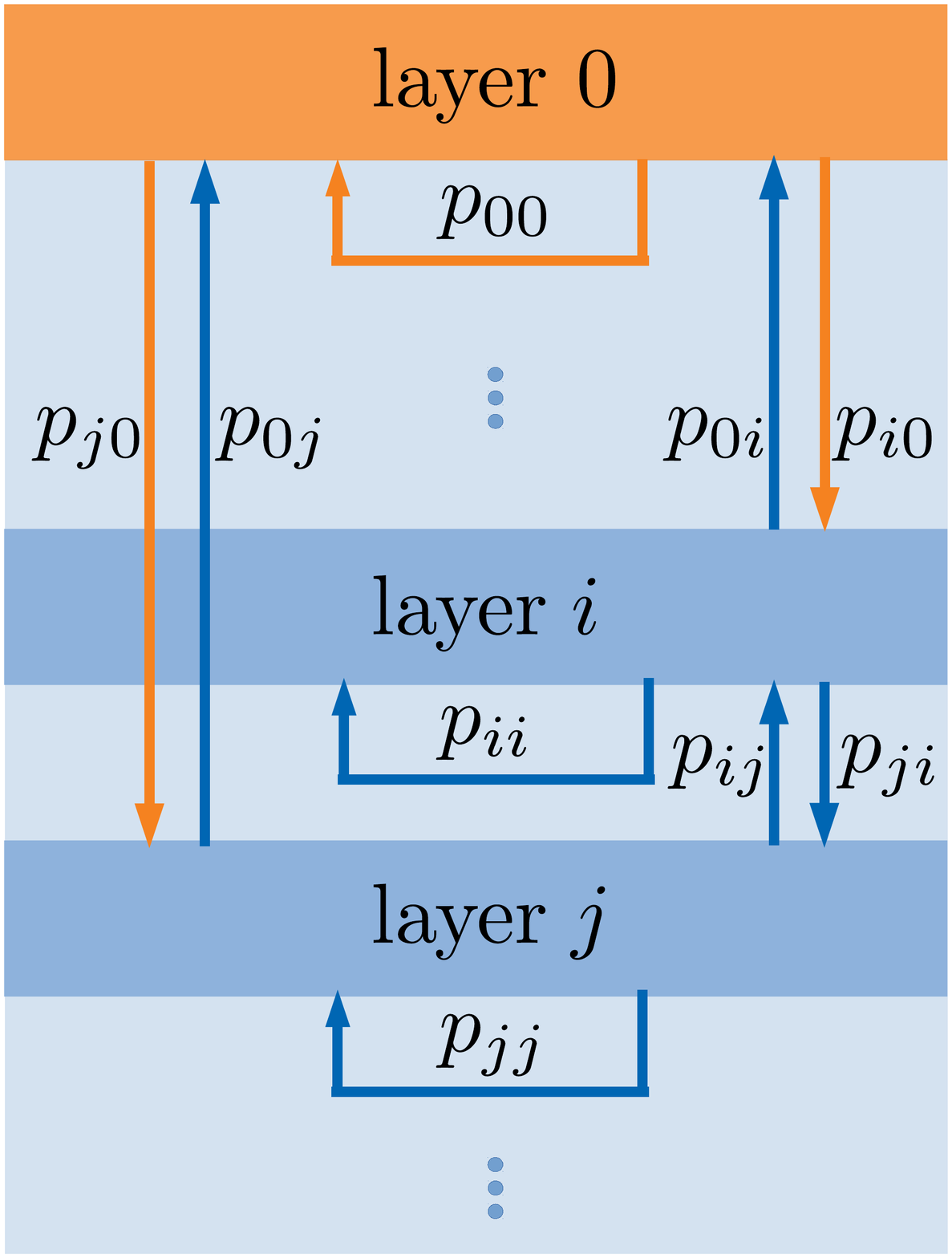}
     }\\
     \end{tabular}

     \caption{(a)  Illustration showing the stratification of the system into layers along with how distances between the layers are computed.  The purple arrow indicates the transition probability density from layer $i$ to layer $j$ at an angle $\theta$: $dp^{(e)}(\theta)p^{(t)}(\theta,i,j)p^{(a)}(\theta)$.  The transition probability $p_{ji}$ is obtained by integrating the transition probability density over $\theta$ from 0 to $\pi/2$.  (b)  A transition diagram for the system.  The rectangles represent the states and the arrows represent the transition probabilities.}
     \label{fig: general diagram and markov graph and transmission distance}
\end{figure}

Under the assumption that emission occurs at the center of layer $i$, the length of the direct path from layer $i$ to layer $j$ is $(j-i-1/2)\Delta z/\cos\theta$ if $i<j$ or $(i-j-1/2)\Delta z/\cos\theta$ if $i>j$, see Fig.~\ref{Markov graph general}.  Thus the direct transmission probability from layer $i$ to layer $j$ along an angle $\theta$ is
\begin{equation}\label{direct transmission probability}
    p^{(t)}_{D}(\theta,i,j)=e^{-\alpha(\theta)(|i-j|-1/2)\Delta z/\cos\theta}.
\end{equation}
(Since $\theta$ is now assumed to be between $0$ and $\pi/2$, we no longer need to take the absolute value of $\cos\theta$).

Calculating $p_R^{(t)}(\theta,i,j)$ involves considering several cases, depending on whether a mirror is present on the bottom layer, and whether $\theta$ is within the escape cone for medium-environment interfaces. For angles within the escape cone $\theta_c$, there is only one possible reflected path, which occurs only if there is a mirror on the bottom layer, and the photon initially travels downward from layer $i$ and is reflected back up to layer $j$. But for angles outside the escape cone, the light can initially travel either upwards or downwards, and can reflect any number of times, regardless of whether the top and bottom are mirrors or interfaces with the environment.

Thus, for $\theta<\theta_c$, the probability of transmission by reflection, $p_R^{(t)}$ is either 0 (if there is no mirror on the bottom) or (in the presence of a mirror) can be calculated by means of the path length of the downwards reflection. This path passes through $M-i+1/2$ layers as it moves downwards (from layer $i$ to the mirror), and then $M-j$ layers after reflection (from the mirror to the bottom of layer $j$). In this case, where the only reflection is a single reflection of a downwards path,
\begin{equation}\label{reflection transmition probability escape cone mirror}
    p_{R}^{\downarrow}(\theta,i,j)=e^{-\alpha(\theta)\Delta z(2M -(i+j)+1/2)/\cos\theta},
\end{equation}

But if $\theta\geq\theta_c$, a photon can travel from $i$ to $j$ by means of any number of reflections of either an upwards or downwards path. It is convenient to write $p_R^{(t)}$ as the sum of several terms: $p^{\downarrow O}_{R}$, representing transmission probability along any path that starts by moving downwards and has an odd number of reflections; $p^{\downarrow E}_{R}$, representing transmission probability along any path that starts by moving downwards and has an even number of reflections; $p^{\uparrow O}_{R}$, representing transmission probability along any path that starts by moving upwards and has an odd number of reflections; and $p^{\uparrow E}_{R}$, representing transmission probability along any path that starts by moving upwards and has an even number of reflections.

For each of these terms, there is a shortest path, with either one or two reflections. The longer paths in each term differ in length by $2M\Delta z/\cos\theta$, since they cross the entire thickness of the medium twice to get two more reflections. Thus, the transmission probability along each path is $e^{-2\alpha(\theta)M\Delta z/\cos\theta}$ times the probability along the path with two fewer reflections. The sum of these transmission probabilities is thus a geometric series, whose value can be calculated by multiplying the first term by $1/(1-e^{-2\alpha(\theta)M\Delta z/\cos\theta})$. To find the first term in the series, we calculate the shortest path length starting with an emission in the relevant direction and either one or two reflections. The shortest path for a photon that starts downward and has an odd number of reflections was given in equation (\ref{reflection transmition probability escape cone mirror}), and the others are calculated similarly:
\begin{align}
    p^{\downarrow O}_{R}(\theta,i,j) &=& \frac{e^{-\alpha(\theta)\Delta z(2M-(i+j)+1/2)/\cos\theta}}{1-e^{-2\alpha M\Delta z/\cos\theta}}\label{reflection transmission probability down odd},\\
    p^{\downarrow E}_{R}(\theta,i,j) &=& \frac{e^{-\alpha(\theta)\Delta z(2M+j-i-1/2)/\cos\theta}}{1-e^{-2\alpha M\Delta z/\cos\theta}}\label{reflection transmission probability down even},\\
    p^{\uparrow O}_{R}(\theta,i,j) &=& \frac{e^{-\alpha(\theta)\Delta z(i+j-3/2)/\cos\theta}}{1-e^{-2\alpha(\theta) M\Delta z/\cos\theta}}\label{reflection transmission probability up odd},\\
    p^{\uparrow E}_{R}(\theta,i,j) &=& \frac{e^{-\alpha(\theta)\Delta z(2M+i-j-1/2)/\cos\theta}}{1-e^{-2\alpha(\theta) M\Delta z/\cos\theta}}\label{reflection transmission probability up even}.
\end{align}

We can input these expressions into Eq.~(\ref{total transmission probability}) and see the following. When there is no mirror at the bottom and $\theta<\theta_c$, then $p^{(t)}(\theta,i,j)$ is just given by the term in  Eq.~(\ref{direct transmission probability}). When there is a mirror at the bottom and $\theta<\theta_c$, then $p^{(t)}(\theta,i,j)$ is the sum of the terms in equations (\ref{direct transmission probability}) and (\ref{reflection transmition probability escape cone mirror}). When $\theta\geq\theta_c$, and there is either a mirror or an environmental interface at the bottom, then $p^{(t)}(\theta,i,j)$ is the sum of the terms in Eqs.~(\ref{direct transmission probability}), (\ref{reflection transmission probability down odd}), (\ref{reflection transmission probability down even}), (\ref{reflection transmission probability up odd}), and (\ref{reflection transmission probability up even}).

%\begin{equation}
%    p^{(t)}(\theta,i,j)=e^{-\alpha(\theta)(|i-j|-1/2)\Delta z/\cos\theta}.
%\end{equation}
%When there is a mirror at the bottom, and $\theta<\theta_c$, then
%\begin{align}
    %p^{(t)}(\theta,i,j)=e^{-\alpha(\theta)(|i-j|-1/2)\Delta z/\cos\theta}+e^{-\alpha(\theta)\Delta z(2M -(i+j)+1/2)/\cos\theta}.
%\end{align}
%When $\theta\geq\theta_c$, and there is either a mirror or an environmental interface at the bottom, then
%\begin{align}
    %p^{(t)}(\theta,i,j)=e^{-\alpha(\theta)(|i-j|-1/2)\Delta z/\cos\theta}+\frac{e^{-\alpha(\theta)\Delta z(i+j-3/2)/\cos\theta}+e^{-\alpha(\theta)\Delta z(2M+i-j-1/2)/\cos\theta}+e^{-\alpha(\theta)\Delta z(2M+j-i-1/2)/\cos\theta}+e^{-\alpha(\theta)\Delta z(2M-(i+j)+1/2)/\cos\theta}}{1-e^{-2\alpha(\theta)M\Delta z/\cos\theta}}.
%\end{align}

Note that many of these terms depend on $e^{-\alpha(\theta)M\Delta z/\cos\theta}$. This is the transmission probability of a photon going through the entire thickness of the medium at angle $\theta$. When the medium is thick and highly absorbing at angle $\theta$, this probability is very close to 0. In this case it is a good approximation to treat the denominator of the fraction as 1 (that is, to take just the first term in these infinite series of reflections, rather than the sum of all terms). It is also a good approximation to ignore the even terms entirely (since even two reflections involves passing completely through the thickness of the medium). Additionally, the odd reflection terms can largely be ignored except for the $\uparrow O$ term when $i$ and $j$ are both small, or the $\downarrow O$ term when $i$ and $j$ are both close to $M$.

\subsubsection{Self-absorption}\label{methods-model self-absorption}

In Secs.~\ref{methods-model system-absorption probability} and \ref{methods-model system-transmission probability}, we made the assumption that $i\neq j$. This was essential because it allowed us to multiply a single absorption probability by the sum of the transmission probabilities along all paths. But when $i=j$, the probability of absorption on the direct path is different from the probability of absorption on reflected paths. On the direct path, a photon emitted from the center of the layer only has half of the thickness of the layer in which it can be absorbed, instead of the full thickness of the layer, though it can be emitted in the upward or downward direction on this angle. But on the reflected paths, the photon can be absorbed anywhere along the full thickness of the layer.

To find the probability of self-absorption on the direct path, $p^{(a)}_S(\theta)$, we use the Beer-Lambert law for a path length of $(1/2)\Delta z/\cos\theta$. We double this probability, to account for both upward and downward paths at angle $\theta$:
\begin{equation}\label{self-absorption probability}
    p^{(a)}_{S}(\theta)= 2(1-e^{-\alpha(\theta) \Delta z/2\cos\theta}).
\end{equation}

Thus, the transition matrix element $p_{ii}$ is given by
\begin{equation}
    p_{ii} = \int_{0}^{\pi/2}dp^{(e)}(\theta)(p^{(a)}_{S}(\theta)+p^{(t)}_{R}(\theta,i,i)p^{(a)}(\theta))\label{self-absorption matrix element},
\end{equation}
where $p^{(a)}(\theta)$ and $p_R^{(t)}(\theta,i,i)$ are as described in Secs.~\ref{methods-model system-absorption probability} and \ref{methods-model system-transmission probability}, respectively. (Recall that $p^{(t)}_R(\theta,i,i)$ is 0 when $\theta<\theta_c$ and there is no mirror on the bottom; it is given by the term in Eq.~(\ref{reflection transmition probability escape cone mirror}) when $\theta<\theta_c$ and there is a mirror on the bottom; and it is given by the sum of the terms in Eqs.~(\ref{reflection transmission probability down odd}), (\ref{reflection transmission probability down even}), (\ref{reflection transmission probability up odd}), and (\ref{reflection transmission probability up even}) when $\theta\geq\theta_c$.)

\subsubsection{Transitions from Environment to Medium}\label{methods-model environment to medium}

In all systems we consider, photons enter the system from a source above layer 1, with some characteristic distribution for that system. Thus, we model transitions from the ``environment'' layer to the medium by
\begin{equation}
    p_{j0} = \int_{0}^{\theta_c} dp_{env}^{(e)}(\theta)p^{(t)}(\theta,0,j)p^{(a)}(\theta).\label{environment to medium transmission}
\end{equation}
Although the integration goes up to the critical angle, the emission density may be zero for some range of angles less than the critical angle.  Specifically, if light from the environment impinges on the medium at an angle $\theta_{e}$ then $dp^{(e)}$ is zero in the range $\theta_{m}<\theta\leq\theta_{c}$.  Here $\theta_{m}=\sin^{-1}([n_{e}/n_{m}]\sin(\theta_{e}))$.  The amount of light, or the emission density $dp^{(e)}_{env}$, from the environment will differ in the different systems we consider, and will thus be described in each of the relevant sections. $p^{(a)}(\theta)$ is the same as in Sec.~\ref{methods-model system-absorption probability}.

However, $p^{(t)}(\theta,0,j)$ has some more significant differences from $p^{(t)}(\theta,i,j)$. Light from the environment is incident precisely at the border of the medium, rather than in the middle of a layer, so the distance traversed is an integer number of layers, rather than a half integer. Furthermore, light from the environment only moves in the downward direction.

Transmission probability from the environment to layer $j$ along the direct path is given by
\begin{equation}\label{direct transmission probability environment}
    p^{(t)}_{D}(\theta,0,j)=e^{-\alpha(\theta)j\Delta z/\cos\theta}.
\end{equation}
For $\theta$ within the escape cone, with no mirror on the bottom, the direct term is the only term, so $p^{(t)}(\theta,0,j)=p^{(t)}_{D}(\theta,0,j).$
For $\theta$ within the escape cone, in cases with a mirror on the bottom, $p^{(t)}(\theta,0,j)=p^{(t)}_{D}(\theta,0,j)+p_{R}^{\downarrow}(\theta,0,j),$ where
\begin{equation}\label{reflection transmition probability escape cone mirror environment}
    p_{R}^{\downarrow}(\theta,0,j)=e^{-\alpha(\theta)\Delta z(2M-j)/\cos\theta},
\end{equation}
Light from outside the system only enters at angles within the escape cone, so we do not consider transmission probabilities at angles outside the escape cone.
%For $\theta$ outside the escape cone, with or without a mirror on the bottom, $p^{(t)}(\theta,0,j)=p^{(t)}_{D}(\theta,0,j)+p_{R}^{\downarrow O}(\theta,0,j)+p_{R}^{\downarrow E}(\theta,0,j),$ where
%\begin{align}
    %p^{\downarrow O}_{R}(\theta,0,j) &=& \frac{e^{-\alpha(\theta)\Delta z(2M-j)/\cos\theta}}{1-e^{-2\alpha M\Delta z/\cos\theta}}\label{reflection transmission probability down odd environment},\\
    %p^{\downarrow E}_{R}(\theta,0,j) &=& \frac{e^{-\alpha(\theta)\Delta z(2M+j-1)/\cos\theta}}{1-e^{-2\alpha M\Delta z/\cos\theta}}\label{reflection transmission probability down even environment}
%\end{align}

\subsubsection{Transitions to Environment}\label{methods-model to environment}

Photons that escape the system are said to transition to the ``environment'' layer. While other transitions can happen at any angle from 0 to $\pi/2$, these transitions can only happen at angles within the escape cone. Thus,
\begin{equation}
    p_{0i} = \int_{0}^{\theta_c} dp^{(e)}(\theta)p^{(t)}(\theta,i,0).\label{transmission to environment}
\end{equation}
$dp^{(e)}$ is just as in Sec.~\ref{methods-model system-emission probability}. We no longer need to account for absorption, because photons that reach the interface with the environment within the escape cone automatically escape.

When there is \emph{no mirror} on the bottom layer, there are two direct paths from layer $i$ to the environment --- an upward and a downward path. Thus, 
$p^{(t)}(\theta,i,0)=p^{\uparrow}(\theta,i,0)+p^{\downarrow}(\theta,i,0),$
where
$p^{\uparrow}(\theta,i,0)=e^{-\alpha(\theta)(i+1/2)\Delta z/\cos\theta}$
and
$p^{\downarrow}(\theta,i,0)=e^{-\alpha(\theta)(M-i+1/2)\Delta z/\cos\theta}.$

When there is \emph{a mirror} on the bottom layer, there is a direct upwards path from layer $i$ to the environment, and a downward path with one reflection. Thus,
$p^{(t)}(\theta,i,0)=p^{\uparrow}(\theta,i,0)+p_{R}^{\downarrow}(\theta,i,0),$
where $p^{\uparrow}(\theta,i,0)$ is as above, and 
$p^{\downarrow}_R(\theta,i,0)=e^{-\alpha(\theta)(2M-i+1/2)\Delta z/\cos\theta}.$

Finally, we must consider transitions directly from the environment to itself.
\begin{equation}
    p_{00} = \int_{0}^{\theta_c} dp^{(e)}_{env}(\theta)p^{(t)}(\theta,0,0).\label{transmission env to env}
\end{equation}
As above, we only consider angles within the escape cone, and there is no need for an absorption term. As in Sec.~\ref{methods-model environment to medium}, $dp^{(e)}_{env}(\theta)$ depends on the system.  As mentioned in Sec.~\ref{methods-model environment to medium} $dp^{(e)}$ may be zero for some angles.

$p^{(t)}(\theta,0,0)$ consists only of a single term, and is either $e^{-\alpha(\theta)2M\Delta z/\cos\theta}$ or $e^{-\alpha(\theta)M\Delta z/\cos\theta}$, depending on whether there is or is not a mirror at the bottom of the medium.

\subsubsection{Variable Quantum Yield}\label{methods-model system-nonunity quantum yield}

So far, we have assumed that every photon that is absorbed by a layer is re-emitted by that layer. That is, we have assumed there is no non-radiative loss, and in particular that the quantum yield of the fluorophores $\eta_Q=1$. However, when there are such losses, these can be modeled as just another sort of transition from each layer of the medium to the dummy ``environment'' layer. We will consider non-radiative losses due to non-unity quantum yield of the fluorophores. %All other transitions are scaled down by a factor of $\eta_Q$.
%The transition matrix discussed until this point does not take into account any non-radiative loss.  We use the term non-radiative loss to primarily describe non-unity quantum yield $\eta_{Q}$.  In modeling non-radiative losses, we assume that the transition probability of the medium is scaled by the quantum yield $\eta_{Q}$.  Scaling down the (emission) probabilities of the medium layers results in a loss of photons.  We have to compensate for diminished probabilities by ensuring each column of the transition matrix remains stochastic, i.e. sums to unity.  This is done by scaling up the emission probability of the environment by exactly the amount that was taken from that of the medium.  We can say that this compensation imposes conservation of the total number of photons.  We can imagine that photon energy that is lost due to non-unity quantum yield makes its way back to the environment where it is again converted back to photons, so that at any time we have the a constant number of photons in the system.  Note that each quantum yield gives rise to a new system, where in each system the total number of photons is conserved.  However, it is not necessarily that the number of photons of one system is equal to that in another.    

To transform the transition matrix $\bm{P}$ assuming unity quantum yield to the matrix $\bm{P'}$ accounting for non-unity quantum yield, we need to slightly modify the transitions. All transition probabilities due to emission of photons from the medium are scaled down by the quantum yield, while the remaining probability is added to the probability of transition to the ``environment''. Transitions from the environment are unchanged.
%Taking the quantum yield into account, we make a simple transformation to the transition matrix $\bm{P}\rightarrow\bm{P}'$.  Implicit in the construction of the transition matrix $\bm{P}$ is unity quantum yield.  As such, the elements of $\bm{P}'$ illustrates the conversion from unity quantum yield to a variable quantum yield.  The transformation from $\bm{P}\rightarrow\bm{P}'$ is given by
\begin{align}
    & p_{ji}' = \eta_{Q}p_{ji} \qquad\qquad\quad\text{$i,j\neq0$}\label{non-radiative loss medium transition matrix element in terms of QY alternate}\\
    & p_{0i}' = \eta_{Q}p_{0i}+(1-\eta_{Q})\quad\text{$i\neq0$}\label{non-radiative loss environment transition matrix element in terms of QY alternate}\\
    & p_{j0}' = p_{j0}\label{no change to first column}
\end{align}

%We have not yet discussed anything specific about about how the environment layer emits or absorbs.  But that does not prevent us from discussing the transformation given above.  Notice that the $p_{j0}$, transitions from the environment to layers of the medium, remain unchanged. The quantum yield does not scale down the emission of the environment layer, it only affects the emission probability of the fluorophores of the medium.  It can be easily verified that each column of $\bm{P}'$ sums to unity.  As such, 
The transformation described by Eqs.~(\ref{non-radiative loss medium transition matrix element in terms of QY alternate}), (\ref{non-radiative loss environment transition matrix element in terms of QY alternate}) and (\ref{no change to first column}) preserves the stochastic nature of the transition matrix and the total number of photons in the fictitious system will be conserved. Its steady state will still correspond to the steady state of the physical system, in which the sum of all losses (radiative and non-radiative) is equal to the number of photons entering the system.

%\begin{figure}[!ht]
%    \centering
%    \includegraphics[width=0.47\textwidth]{figures/markov_chain_processes_alternative.%pdf}
%    \caption{Transition from layer $i$ to layer $j$.}
%    \label{transition_figure}
%\end{figure}

%\begin{figure}
%    \centering
%    \includegraphics[width=0.45\textwidth]{figures/markov_model_mirror_general.pdf}
%    \caption{Caption}
%    \label{fig:markov model reflections}
%\end{figure}

%\subsubsection{Reversible Markov Chain}
%For non-zero quantum yield, the transition matrix used in this work will yield a steady state distribution.  In the steady state, the \textit{reversibility} property holds:
%\begin{equation}\label{reversibility property}  
%p_{ji}d^{(s)}_{i} = p_{ij}d^{(s)}_{j}.
%\end{equation}
%The reversibility property tells us that that the number of photons transitioning from layer $i$ to layer $j$ (left-hand side of Eq.~(\ref{reversibility property})) is equal to the number of photons transitioning from layer $j$ to layer $i$ (right-hand side of Eq.~(\ref{reversibility property})).  The reversibility property is also called \textit{detailed balance}.

\section{Results and Discussion}\label{results}
\subsection{Deviations from Beer-Lambert's Law}\label{results-beer-lambert law}
 The well-known Beer-Lambert Law allows one to analyze how a non-fluorescent and non-scattering body absorbs and transmits radiation. A standard experimental geometry usually consists of a collimated light source incident on a sample with known thickness, so the path length of the light as well as the incident and transmitted intensity can be quantified. From this, one can determine the absorption coefficient of a sample, or if the absorption coefficient is known, one can predict the transmitted intensity.  
 
 However, when the sample is strongly fluorescent and/or scatters light at the same wavelength as it absorbs or emits, it is not straightforward to analyze the optical response. Photons can be absorbed, re-emitted, and/or scattered multiple times before exiting the material, and the overall behavior is not well described by the Beer-Lambert Law. By explicitly accounting for these more complex interactions, our model can predict the transmitted intensity; angle-dependent emission and scattering profile into the environment; and the light intensity, i.e. photon population, as a function of position in the medium. The latter information can be used to calculate local variations in the photo-induced chemical potential.  

For analyzing these deviations from Beer-Lambert's law we consider two systems: one is a slab of the photoluminescent semiconducting material gallium arsenide (GaAs) and the other is a colloidal dispersion of gold nanoparticles with strong optical scattering.  In both systems, the medium is sandwiched between two environment layers (no mirror at the bottom), see Fig.~\ref{beer-lambert gallium arsenide} and Fig.~\ref{beer-lambert gold NPs}.  Monochromatic light from the environment impinges upon the media at normal incidence i.e. $\theta_{e}=0$.  %For the slab, we consider emission and absorption.  For the colloid, instead of emission we consider elastic scattering and instead of absorption, we consider extinction.  In both cases we vary $\eta_{Q}$, this parameter was given the moniker fluorescence quantum yield (which holds in the case of emission).  In the case of scattering, $\eta_{Q}$ represents the albedo of scattering, which we label as $\tilde{\omega}$ for clarity.  In both cases, we assume the media is thick enough to absorb \textcolor{red}{$99.xx$\% of the incident light}.

%As mentioned in Sec.~\ref{methods-model environment to medium}, we treat the emission from the environment on a case-by-case basis.  For the case at hand we assume the emission from the environment occurs at one angle, namely $\theta_{e}=0$.  
%Since we are assuming that light from the environment is coming from one angle (collimated) we have  $p^{(e)}_{env}(0)=1$, i.e. all the light from the environment enters the medium at this angle.  %$\theta_{m}$ is the angle at which light enters the medium due to refraction, here, $\theta_{e}=\theta_{m}=0$ (normal incidence).  For oblique incidence, $\theta_{m}$ can be obtained by Snell's law for a given $n_{m}$. % This would be a place to mention that the azimuthal symmetry means this is the same as coming in from a cone.

Because light from the environment is received only at a single angle, the probability \emph{density} is given by the Dirac delta function $dp^{(e)}_{env}(\theta)=\delta(\theta)$. Thus, the integrals in Secs.~\ref{methods-model environment to medium} and \ref{methods-model to environment} can be simplified. Note that light from the environment only travels at angle $\theta=0$ and
there is no mirror on the bottom, so the transmission probabilities only have a single term, yielding
\begin{equation}
    p_{j0} = e^{-\alpha(0)j\Delta z}p^{(a)}(0),\label{beer-lambert environment to medium transmission}
\end{equation}
and
\begin{equation}
    p_{00} = e^{-\alpha(0)M\Delta z}.
\end{equation}

\subsubsection{Gallium Arsenide Slab}\label{GaAs Beer Lambert}

\begin{figure}[!ht]
\centering
    \begin{tabular}[t]{r}
    %\hline
     \subfloat[\label{beer-lambert gallium arsenide}]{%
       \includegraphics[height=0.205\textwidth,width=0.135\textwidth]{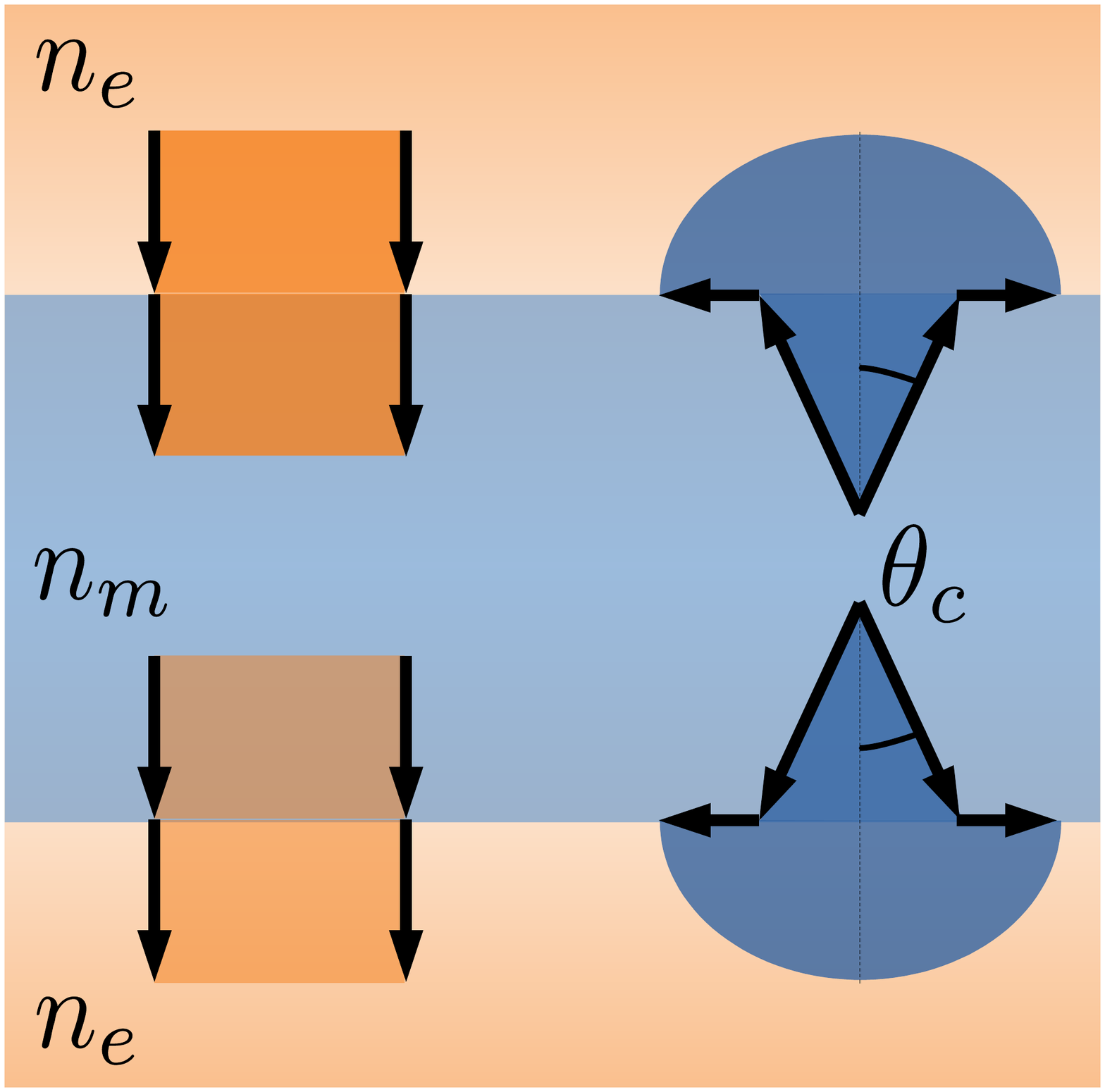}
     }
%     \\
%\hline
     \subfloat[\label{beer-lambert GaAs plot}]{%
      \includegraphics[width=0.35\textwidth]{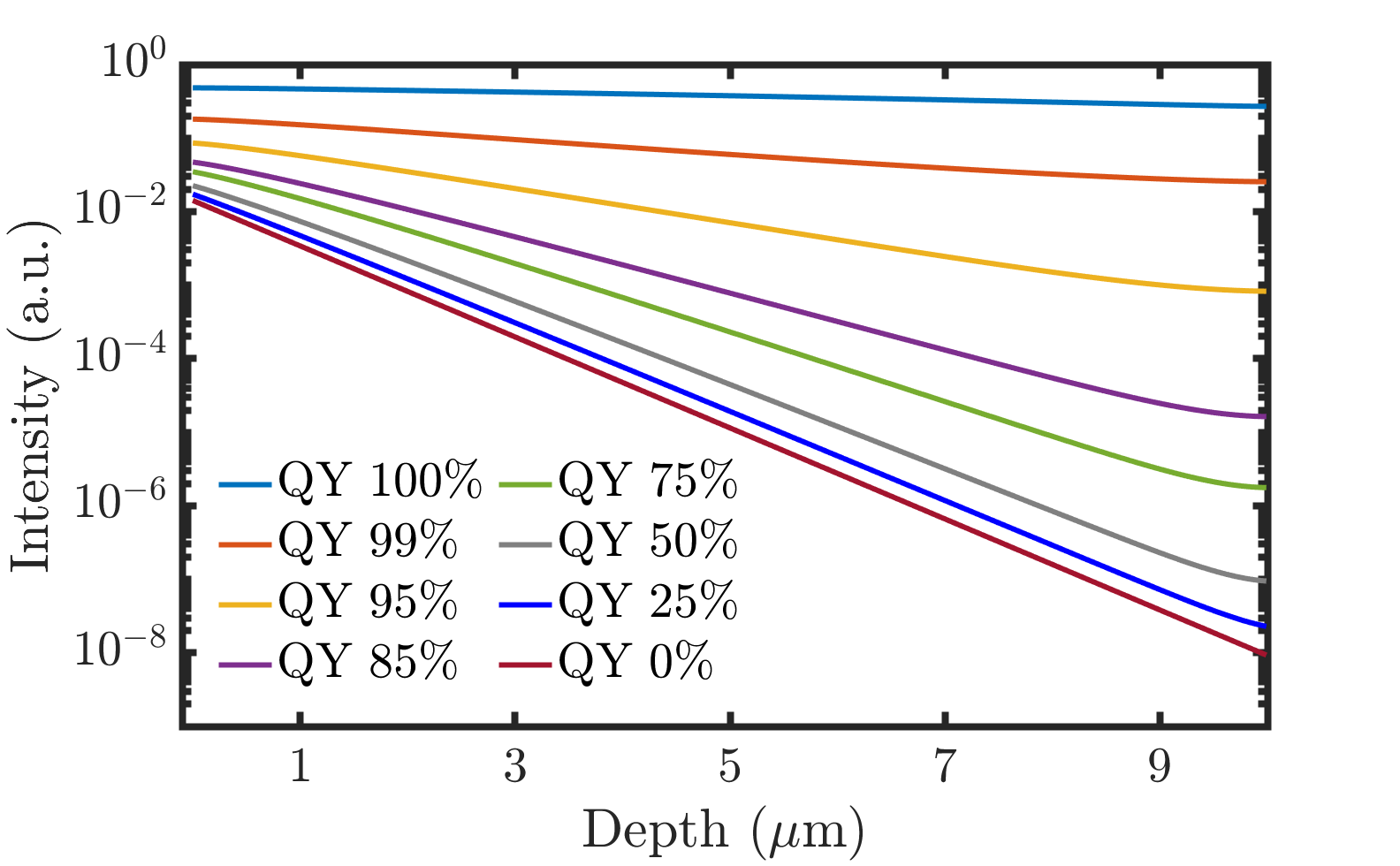}
     }\\
     \end{tabular}

     \caption{(a)  The illustration depicts incoming light in orange impinging at normal incidence.  The light emitted by the medium is in dark blue.  The environment is represented by light orange and the medium by light blue.  (b)  This log-scaled plot shows the steady state photon population as a function of depth for various fluorescent quantum yields in a 10 $\mu$m thick slab of GaAs illuminated at its band gap energy at normal incidence.}
     \label{fig:beer-Lambert for GaAs slab}
\end{figure}

%Now that we have the transition probabilities, let us discuss the GaAs slab first. 
We model GaAs as an isotropic emitter.  We take the absorption coefficient and the refractive index of GaAs at a wavelength corresponding to its bandgap energy at room temperature, i.e. $\alpha = 14247$ cm$^{-1}$, and $n_{m}=3.63$.  The GaAs slab is $10^{-3}$ cm (or 10 microns) thick, as such the entire slab absorbs $99.99994\%$ of light impinging upon it at normal incidence, assuming the light is monochromatic. %We simulate this system with fluorescence quantum yield $\eta_{Q}$ at values ranging from zero to one. 
Fig.~\ref{beer-lambert GaAs plot} plots the steady state population as a function of depth for different cases that analyze how its behavior depends on changes in quantum yield $\eta_Q$ with the incident light intensity held constant.

The case of $\eta_{Q}=0$ corresponds to when Beer-Lambert's law should be strictly valid (no emission, only absorption). %, where one typically measures how absorbing a non-emitting medium is by comparing the incident intensity of a source beam to the transmitted intensity.  Setting $\eta_{Q}=0$ in the transition matrix corresponds to no emission at all from the GaAs slab, so light from the environment is absorbed without being re-emitted.
This case is given by the last trace in Fig.~\ref{beer-lambert GaAs plot}. The constant negative slope corresponds to exponential decay, consistent with usual Beer-Lambert behavior.
%, one observes a constant negative slope (which would be a negative exponential on a linear vertical axis), this result is expected for the usual Beer-Lambert's law.  

%\begin{figure}
%    \centering
%    \includegraphics[width=0.45\textwidth]{figures/beer_law_GaAs_small_illustration_and_plot_v2.pdf}
%    \caption{\textbf{Left:} The illustration depicts incoming light in orange impinging at normal incidence.  The light emitted by the medium is in dark blue.  The environment is represented by light orange and the medium by light blue.  \textbf{Right:} The plot shows the intensity as a function of depth for various fluorescent quantum yields.}
%    \label{fig:beer-Lambert for GaAs slab}
%\end{figure}

%We saw for the case of $\eta_{Q}=0$, the steady state distribution decayed exponentially as a function of depth (or increasing number of layers).  However, 
Deviations from the Beer-Lambert law occur when emission is ``turned on", i.e. $\eta_{Q}\neq 0$.  As $\eta_{Q}$ increases, the steady state population of each layer increases monotonically. At any $\eta_Q$, population decreases with depth --- it is always higher towards the top, where light is incident, and lower towards the bottom, where light is only emitted to the environment. The decrease of population with depth is less pronounced for higher $\eta_Q$, but even with $\eta_Q=1$, the steady state distribution is not uniform. Most significantly, the Markov model enables us to calculate the  deviation from strict exponential decay for any $\eta_Q>0$.  Such deviations can be clearly seen in Fig.~\ref{fig:beer-Lambert for GaAs slab}.  Upon close inspection, one observes a subtle tilde-shaped features to plots in Fig.~\ref{fig:beer-Lambert for GaAs slab} especially for $0.5\leq\eta_{Q}\leq 0.99$.  These features are due to total internal reflections.  The internally reflected light, reflecting off the top and bottom interfaces has a higher probability of being absorbed in the layers nearest to the interface where the photons were reflected.  This causes an increase in the population near the top and bottom of the medium leading the to curvy features observed in Fig.~\ref{fig:beer-Lambert for GaAs slab}. 
%It is interesting to note that we do not have a perfectly constant steady state distribution for unity quantum yield $\eta_{Q}=1$.  Unity quantum yield corresponds to each absorbed photon absorbed being re-emitted without loss.  We do not have a constant steady state distribution for the case of unity quantum yield because of our boundary conditions.  That is, we assume that the medium is illuminated from the top only, but can escape from both the top and bottom of the medium. %(interface 0), and light can escape from both the top (interface 0) and the bottom (interface $M$) of the medium.  In particular, layers near interface 0 of the medium have higher probabilities of absorbing photons originating from interface 0 than layers of the medium near interface $M$.  While photons emitted from layers near interface $0$ and $M$ have similar probabilities of escaping the medium, say for example the third layer from interface 0 has the same transition probability of its photons escaping the medium at interface 0 as the photons emitted form third layer from interface $M$.  In other words, the transmission probability from layer 0 to layer 3 is greater than that of the the transmission probability from layer 0 to layer $M-2$ (third layer from interface $M$), this transmission probability governs the absorption of photons from the environment to said layer.  However, it is equally likely that said layers emit photons to the environment since the distance from interface 0 to layer 3 is equal to the distance from interface $M$ to layer $M-2$. 

\subsubsection{Colloidal Gold Nanoparticles}\label{subsec colloidal gold}

%Now let us examine the case of scattering with no emission.  
We consider a colloid composed of gold nanospheres in water. For this model, the ``absorption'' and ``fluorescence'' terms developed above are interpreted to represent incoming and outgoing light, respectively, that is scattered by the nanoparticles.
As a simple approximation we assume that the gold spheres scatter light isotropically. A more realistic model would involve Rayleigh-Gans or Lorenz-Mie \emph{phase functions}; the scattering phase function specifies the angular distribution of the scattered light \cite{Lenoble2013}. Since the angular distribution of scattered light depends on the incident angle, a Markov model for these systems would need to track not just the location of the most recent interaction, but also the incident angle, in its state space. Here we use isotropic scattering as a first rough approximation that can be calculated with the simpler state space.
Models where the phase function is employed for light scattering using Markov chains are considered in Refs.\cite{Esposito1978,Li2016,Lin2016}.

For comparison with common experimental conditions, the number density of the gold particles is $7.15\times 10^{15}$ m$^{-3}$, the extinction cross section is $2.93\times 10^{-15}$ m$^{2}$.  The thickness of the medium is $1$ cm and the refractive index is taken to be that of water 1.33.  For our analysis we consider behavior that results when the extinction cross-section $\sigma_{e}$, which is the sum of the absorption cross-section $\sigma$ and the scattering cross-section $\sigma_{s}$, remains fixed as the absorption and scattering cross-section are varied.  We keep track of only the variation of the scattering cross-section in terms of the extinction cross-section $\tilde{\omega}=\sigma_{s}/\sigma_{e}$, where $\tilde{\omega}$ is called the scattering albedo.  The  scattering albedo is varied from zero to one.  For a scattering albedo of zero the particles only absorb light (no scattering), whereas for a scattering albedo of one, the particle only scatters light (no absorption).

We assume that the particles and incident light satisfy the conditions required for elastic and isotropic scattering.  %We can use the same formalism used to describe emission above, except that here we interpret the emission probability density as a scattering probability density and the quantum yield as a scattering albedo.%To change the scattering albedo we assume the size of the particle varies as well.  

%\begin{figure}[h!t]
%    \centering
%    \includegraphics[width=0.45\textwidth]{figures/beer-lambert_gold_np_plot_and_small_schematic_v2.pdf}
%    \caption{\textbf{Left:} Illustration of a cuvette filled with colloidal gold nanoparticles.  A laser beam impinges at normal incidence upon the cuvette. \textbf{Right:} Plot of the scattered intensity as a function of depth for various scattering albedos.}
%    \label{fig:beer-Lambert for Au nanoparticles}
%\end{figure}

\begin{figure}[h!t]
\centering
    \begin{tabular}[t]{r}
    %\hline
     \subfloat[\label{beer-lambert gold NPs}]{%
       \includegraphics[height=0.205\textwidth,width=0.135\textwidth]{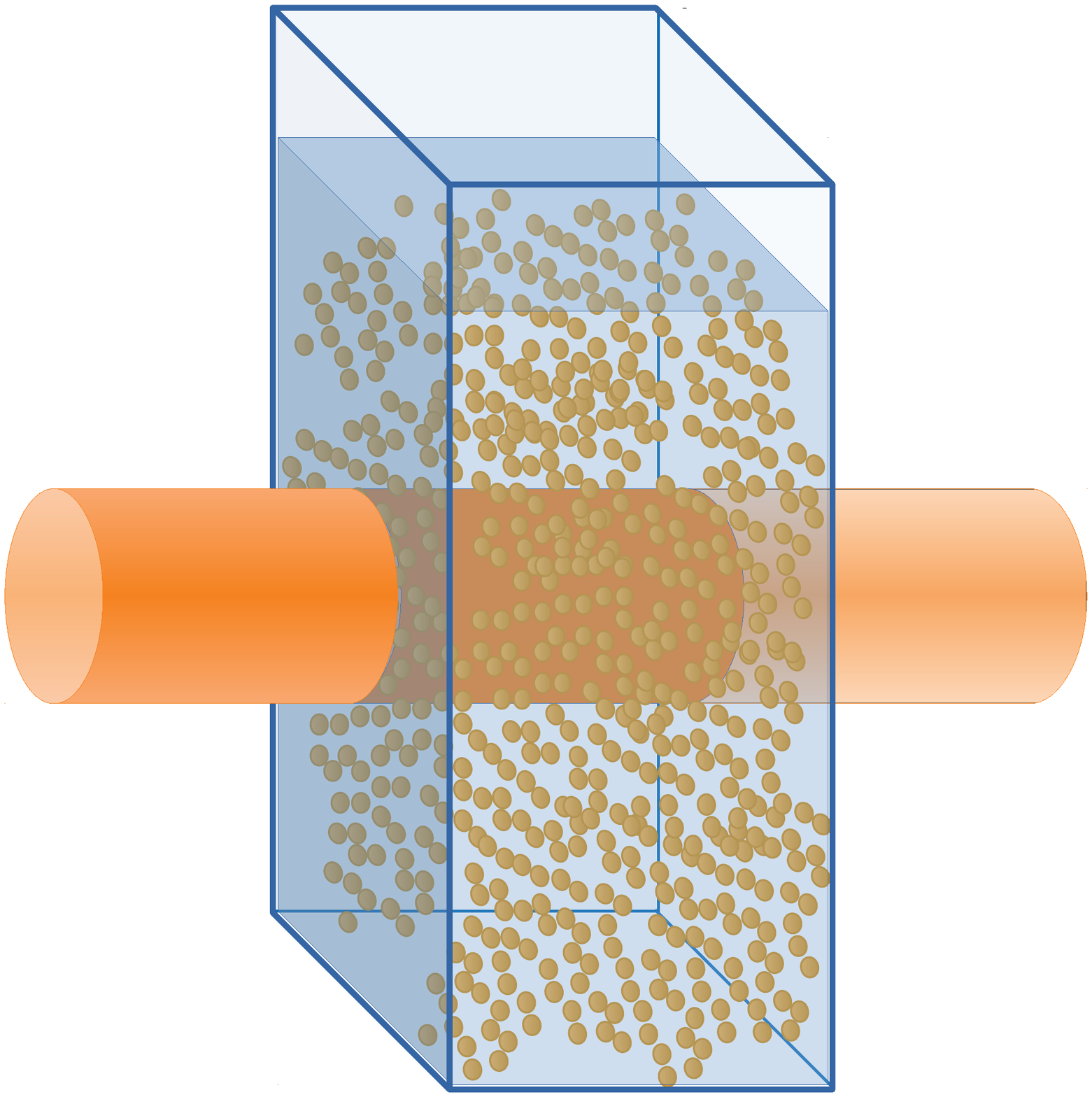}
     }
%     \\
%\hline
     \subfloat[\label{beer-lambert gold NPs plot}]{%
      \includegraphics[width=0.35\textwidth]{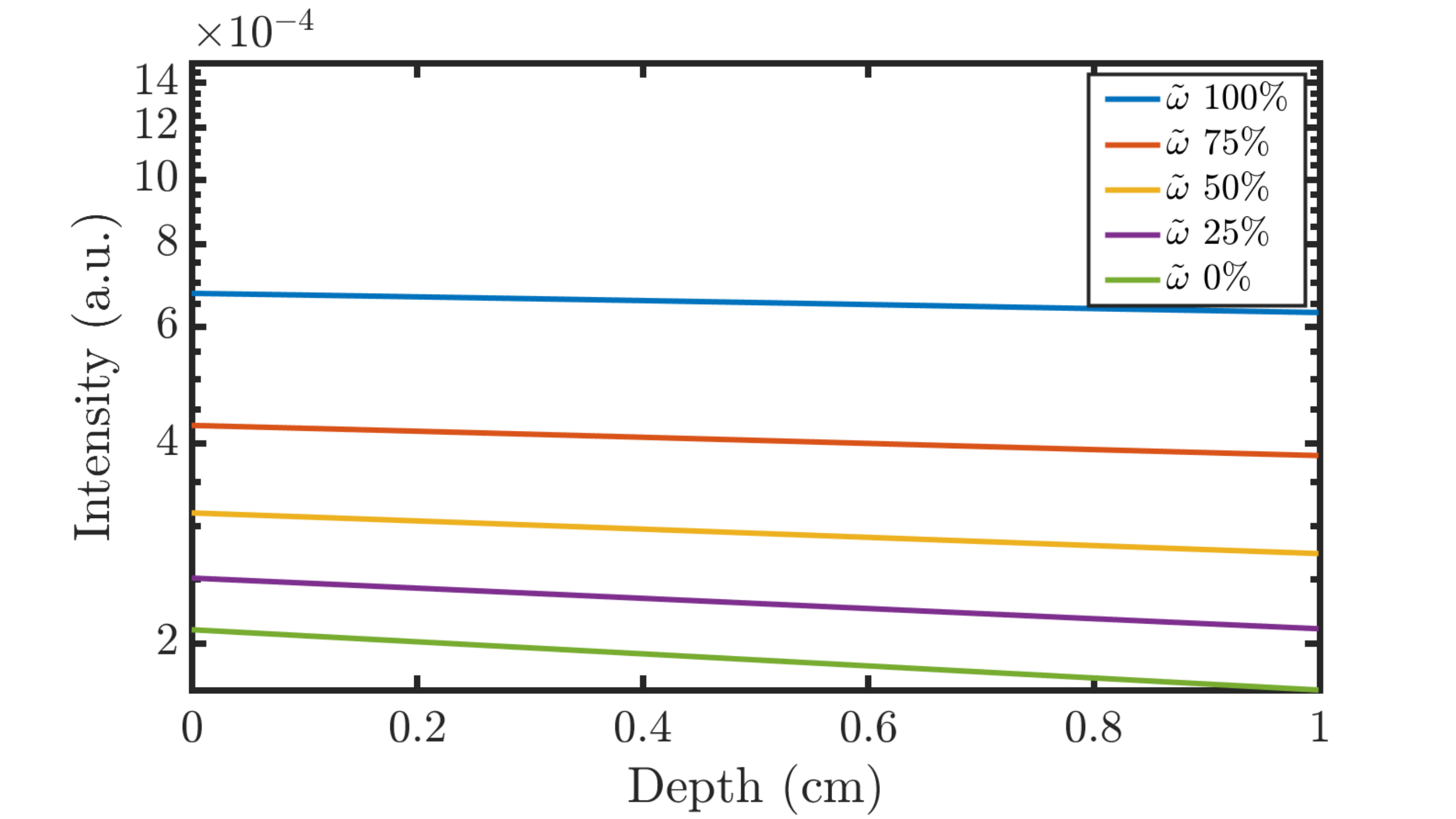}
     }\\
     \end{tabular}

     \caption{(a)  Illustration of a cuvette filled with colloidal gold nanoparticles. The orange cylinder represents collimated light impinging on the cuvette at normal incidence.  (b)  Plot of the light intensity as a function of depth for various scattering albedos, $\tilde{\omega}$.}
     \label{fig:beer-Lambert for Au nanoparticles}
\end{figure}

In the case of zero scattering albedo, one sees from Fig.~\ref{fig:beer-Lambert for Au nanoparticles} that there is an exponential decay of photons (the vertical axis is logarithmic), as expected.  As the scattering albedo increases, the photons have a chance of being scattered.  It is apparent from Fig.~\ref{fig:beer-Lambert for Au nanoparticles} that the attenuation increases as the scattering albedo decreases.

\subsection{Kirchhoff's Law and Violations of Detailed Balance}\label{results kirchhoffs law}

In this section we reproduce Kirchhoff's law of radiation, and analyze deviations from it. Kirchhoff's law of radiation states that the emissivity of a medium at any angle is equal to its absorptivity at that angle, in thermal equilibrium. This is a stricter condition than a steady state, in which it is only required that the total flow of photons from the medium to the environment is equal to the total flow of photons from the environment to the medium. 

Kirchhoff's law requires that in equilibrium there is also a \emph{detailed balance}, meaning that the flow of photons from the medium to the environment \emph{at an angle} is equal to the flow of photons from the environment to the medium at that same angle. We show that although this detailed balance holds in equilibrium systems, and is very well approximated in a non-equilibrium steady state system with isotropic fluorophores, it is clearly violated in a non-equilibrium steady state system with aligned dipolar fluorophores.

We consider two types of environmental radiation interacting with a medium. The first is an isotropic radiation field characteristic of thermal equilibrium, e.g. the interior radiation field of a blackbody cavity. Under these conditions, light is incident on the medium with a Lambertian angular profile \cite{Melvin1955}, with
\begin{equation}\label{incident lambertian}
    dp^{(e)}_{env}(\theta) = \frac{2}{\sin^{2}\theta_{c}}\cos\theta\sin\theta d\theta,
\end{equation}
for $\theta<\theta_c$. The second radiation source we consider is light received from the solar disk, i.e. sunlight, at normal incidence to the medium. We use $dp^{(e)}_{env}(\theta)=\nu\sin\theta \cos\theta d\theta$ for $\theta\leq0.267^{\circ}$ and $dp^{(e)}_{env}(\theta)=0$ for $\theta>0.267^{\circ}$, where $\nu$ is a normalization constant equal to $n_{m}^{2}/(\pi n_{e}^{2}\sin^{2}(0.267^{\circ}))$. Since our medium emits at all angles, but absorbs only at angles corresponding to the solar disk, its steady state is not an equilibrium.

We also consider two versions of the medium, one with isotropic fluorophores and the other with dipolar fluorophores. The absorption, transmission, and emission probabilities for these are defined in Sec.~\ref{methods-model system}. For the isotropic case, we consider a slab of GaAs with parameters (thickness, refractive index, absorption coefficient and emission/absorption wavelength of light) as described in Sec.~\ref{GaAs Beer Lambert}. For the dipolar case, we consider cadmium selenide/cadmium sulfide (CdSe/CdS) dot-in-rod crystals that are in a chloroform solution, and are aligned with their long axis normal to the surface of the medium. The concentration of the CdSe/CdS dot-in-rod fluorophores is $10^{-6}$ molar concentration or molarity (M), extinction coefficient of the fluorophores is $5\times10^{6}$ cm$^{-1}$M$^{-1}$, the wavelength of emission/absorption is taken as 615 nm.  The refractive index of the medium is that of chloroform at 615 nm, i.e. 1.445.  The thickness of the medium is 2 cm.  

In each of the four cases (Lambertian or solar incident radiation; isotropic or dipolar fluorophores) we always have a mirror at the bottom of layer $M$, and assume the refractive index of the environment is $n_e=1$.

%If the incident radiation on the top surface of the medium is Lambertian, and there is a mirror at the bottom of the medium, then the steady state should be an equilibrium. Our model will demonstrate that Kirchhoff's law is upheld in this case regardless of the medium. But if the incident radiation is solar, then the steady state will not be an equilibrium. However, we can still use the model to calculate the angular distribution of emission from the medium and compare it to the incident radiation to see how detailed balance is violated. We observe that when the medium is isotropic, the steady-state behavior is well-approximated by the equiilibrium Lambertian emission, but when the medium is dipolar, the steady-state behavior is quite different.

To probe the angle distribution of photons entering and leaving the system, we partition the range $0\leq\theta\leq\theta_{c}$ into a number $b_{max}$ of bins, each of size $\Delta\theta=\theta_c/b_{max}$. We index these bins by a value $b$ ranging from $1$ to $b_{max}$. The bin $b$ consists of angles from $(b-1)\Delta\theta$ to $b\Delta\theta$.  Note that for the equilibrium case the light entering the medium, upon refraction, spans the range $\theta\leq\theta_{c}$.  However, for the non-equilibrium case, the angular range of light entering the medium is less than $\theta_c$.  In particular, it spans the range $\theta\leq\sin^{-1}([n_{e}/n_{m}]\sin(0.267^{\circ}))$.  However, we still use the same bin range while keeping in mind that the absorption corresponding to bins in the range $\sin^{-1}([n_{e}/n_{m}]\sin(0.267^{\circ}))<\theta\leq\theta_{c}$ is zero. 

We analyze transitions in greater detail now by identifying the transition probabilities at angles within each bin. That is, we break up the definitions of the transition probabilities $p_{0i}$ and $p_{j0}$ so that instead of integrating over all angles, we integrate just over angles within the bin. We will have $p_{0i}=\sum_{b=1}^{b_{max}}p_{0i}(b)$ and $p_{j0}=\sum_{b=1}^{b_{max}}p_{j0}(b)$.
For $i>0$, $p_{0i}(b)$ is the probability of a photon in layer $i$ to leave the medium at an angle in bin $b$:
\begin{equation}
    p_{0i}(b)=\int_{(b-1)\Delta\theta}^{b\Delta\theta} dp^{(e)}(\theta)p^{(t)}(\theta,i,0),\label{angle dependent env to medium}
\end{equation}
as in Eq.~(\ref{transmission to environment}).  Similarly, for $j>0$, $p_{j0}(b)$ is the probability of a photon from the environment to reach layer $j$ at an angle in bin $b$:
\begin{equation}
    p_{j0}(b) = \int_{(b-1)\Delta\theta}^{b\Delta\theta} dp_{env}^{(e)}(\theta)p^{(t)}(\theta,0,j)p^{(a)}(\theta),\label{angle dependent environment to medium transmission}
\end{equation}
as in  Eq.~(\ref{environment to medium transmission}).
$p_{00}(b)$ is the probability of light from the environment entering the medium, being reflected off the mirror on the bottom, and emerging again without being absorbed in the medium. That is:
\begin{equation}
    p_{00}(b) = \int_{(b-1)\Delta\theta}^{b\Delta\theta} dp^{(e)}_{env}(\theta)p^{(t)}(\theta,0,0),\label{angle dependent env to env}
\end{equation}
as in Eq.~(\ref{transmission env to env}).

Once we have these detailed probabilities, we can analyze the total absorption and emission of the medium within each of these bins by summing over all layers. In the steady state, the angle-dependent absorption of the medium at angles in bin $b$ is
\begin{equation}
    a(b)=\sum_{j=0}^{M}d^{(s)}_0 p_{j0}(b),\label{angle-dependent absorption}
\end{equation}
and the angle-dependent emission of the medium at angles in $b$ is
\begin{equation}
    e(b)=\sum_{i=0}^{M}d^{(s)}_ip_{0i}(b).\label{angle-dependent emission}
\end{equation}
Note that $a(b)$ can be simplified to $\int_{(b-1)\Delta\theta}^{b\Delta\theta}dp^{(e)}_{env}(\theta)$, because the sum of $p^{(t)}(\theta,0,j)p^{(a)}(\theta)$ is 1 (since all light must end up somewhere), though there is no comparable simplification of $e(b)$. Because this is the steady state, it is clear that $\sum_{b=1}^{b_{max}}a(b)=\sum_{b=1}^{b_{max}}e(b)$. But the detailed balance predicted by Kirchhoff's law in equilibrium would require that $a(b)=e(b)$ for each bin $b$.

\begin{figure}
\begin{tabular}{cc}
    \subfloat[\label{kirchhoff ill iso}]{%
       \includegraphics[width=0.145\textwidth]{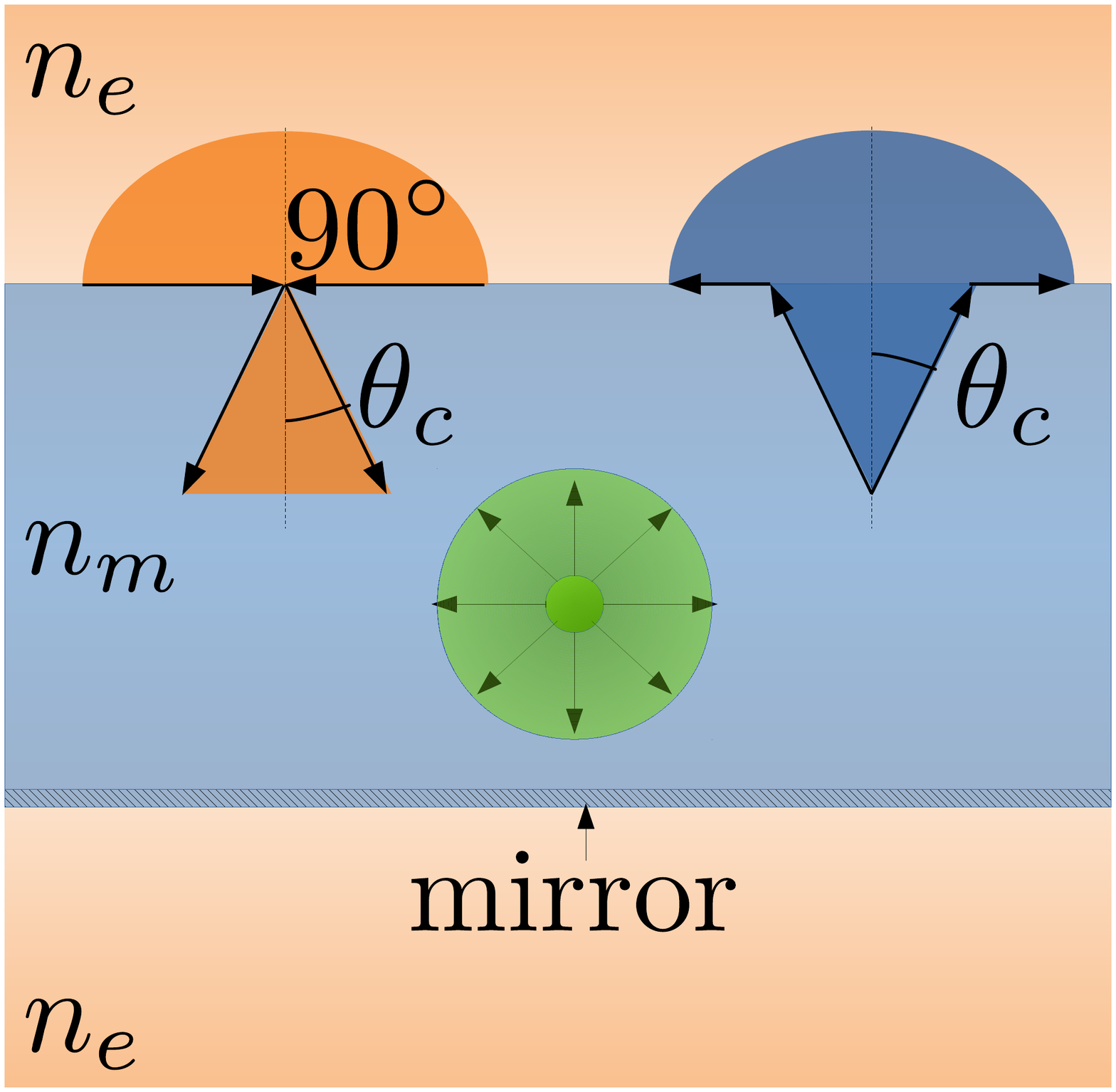}
     } & \subfloat[\label{kirchhoff plot iso}]{%
       \includegraphics[width=0.29\textwidth]{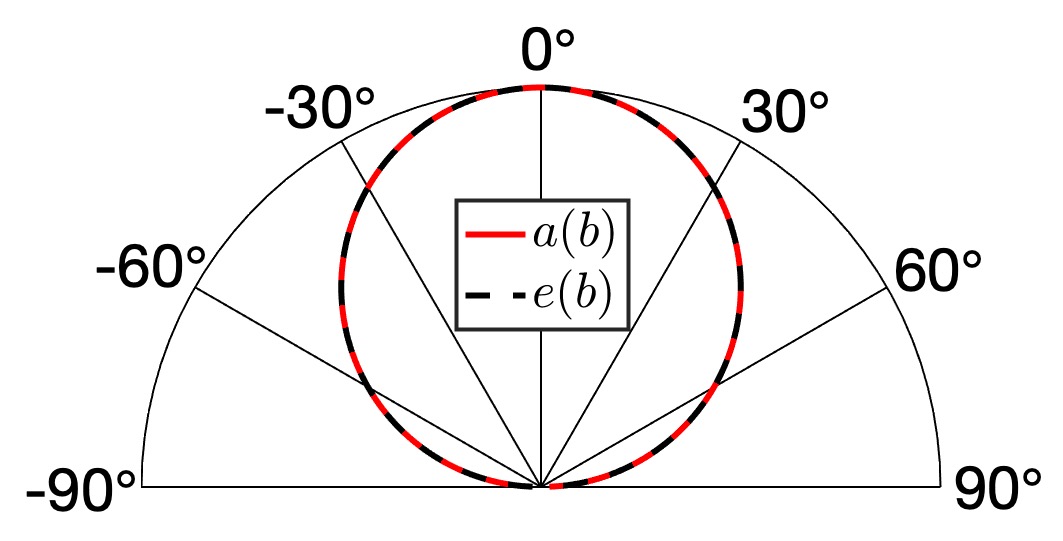}
     } \\
    \subfloat[\label{kirchhoff ill dip}]{%
       \includegraphics[width=0.145\textwidth]{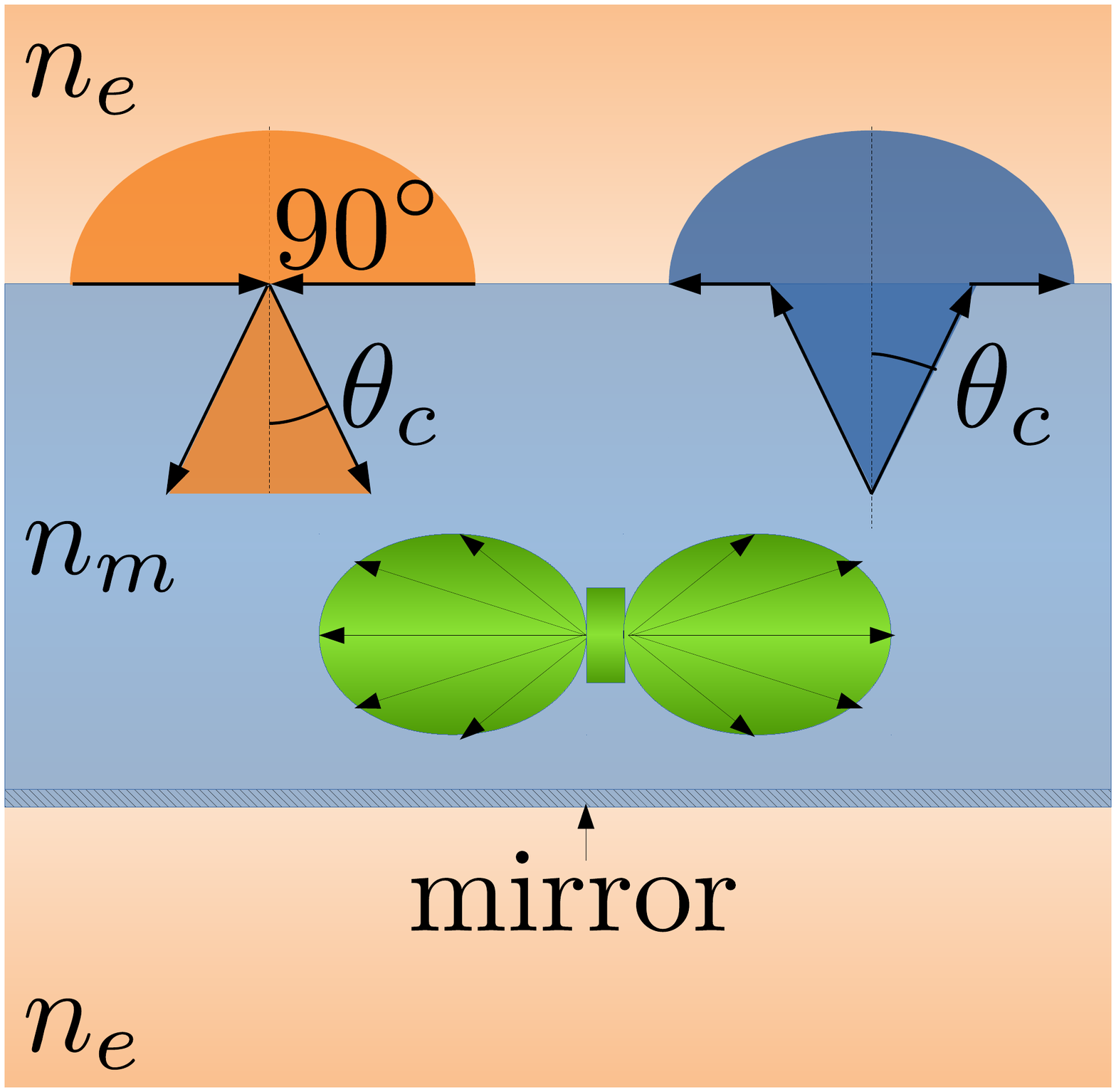}
     } & \subfloat[\label{kirchhoff plot dip}]{%
       \includegraphics[width=0.29\textwidth]{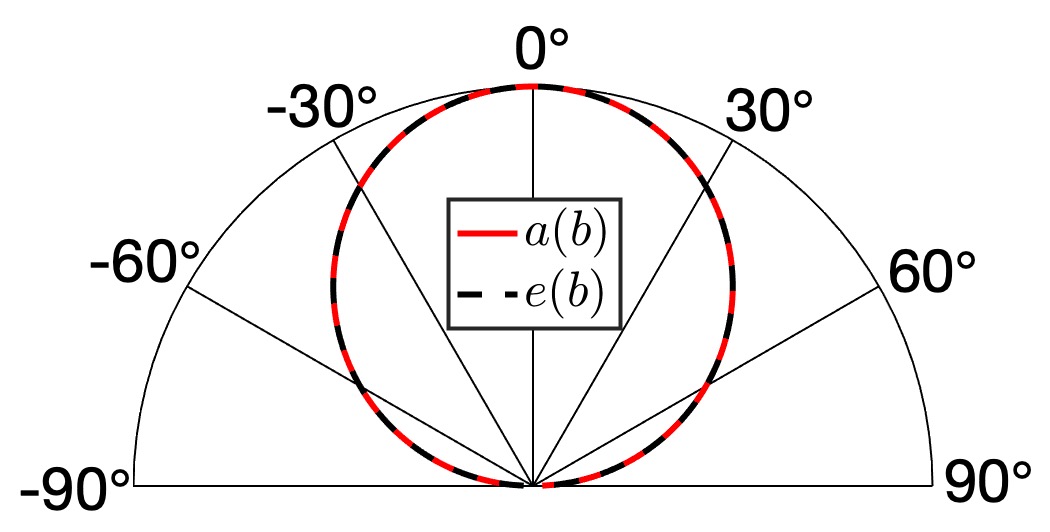}
     } \\
\end{tabular}
\caption{Kirchhoff's Law Analysis.  The equilibrium case. (a) and (c) illustrate the two types of media considered, one that emits isotropically (the green circular object in (a)) and another that emits like a dipole (the green two-lobed object in (c)).  For both (a) and (c) the orange semicircle and orange structures represent the incident light whereas the dark blue represents the light exiting the medium.  (b) is a polar plot of the light entering (red trace) the medium and light leaving the medium (black trace) in the steady state for the isotropic emitting medium.  (d) is the same as (b) except for the dipole emitting case. }
     \label{kirchhoff figure equilibrium}
\end{figure}

%%%%%% nonequilibrium figure %%%%%
\begin{figure}
\begin{tabular}{cc}
    \subfloat[\label{kirchhoff ill iso noneq}]{%
       \includegraphics[width=0.145\textwidth]{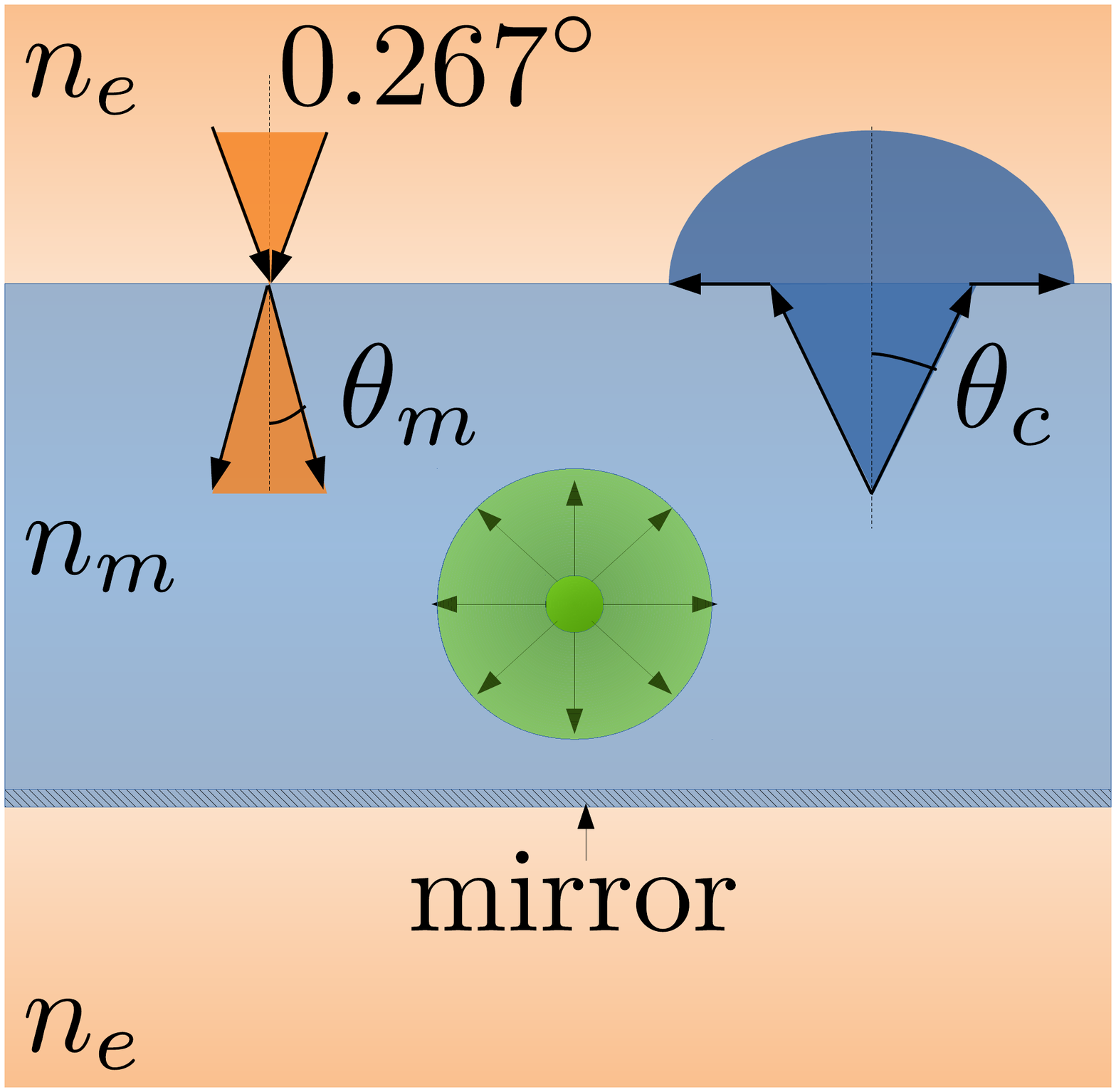}
     } & \subfloat[\label{kirchhoff plot iso noneq}]{%
       \includegraphics[width=0.29\textwidth]{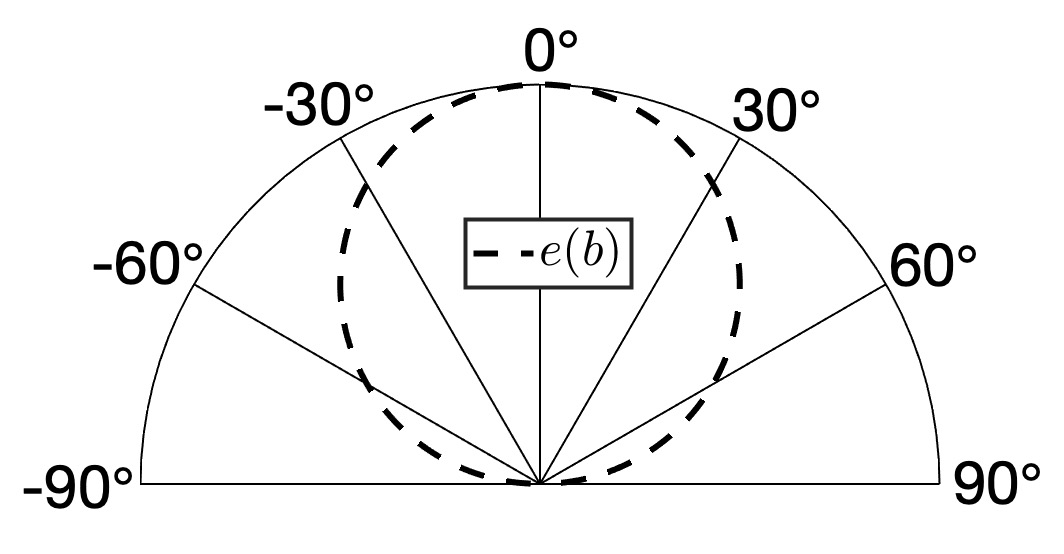}
     } \\
    \subfloat[\label{kirchhoff ill dip noneq}]{%
       \includegraphics[width=0.145\textwidth]{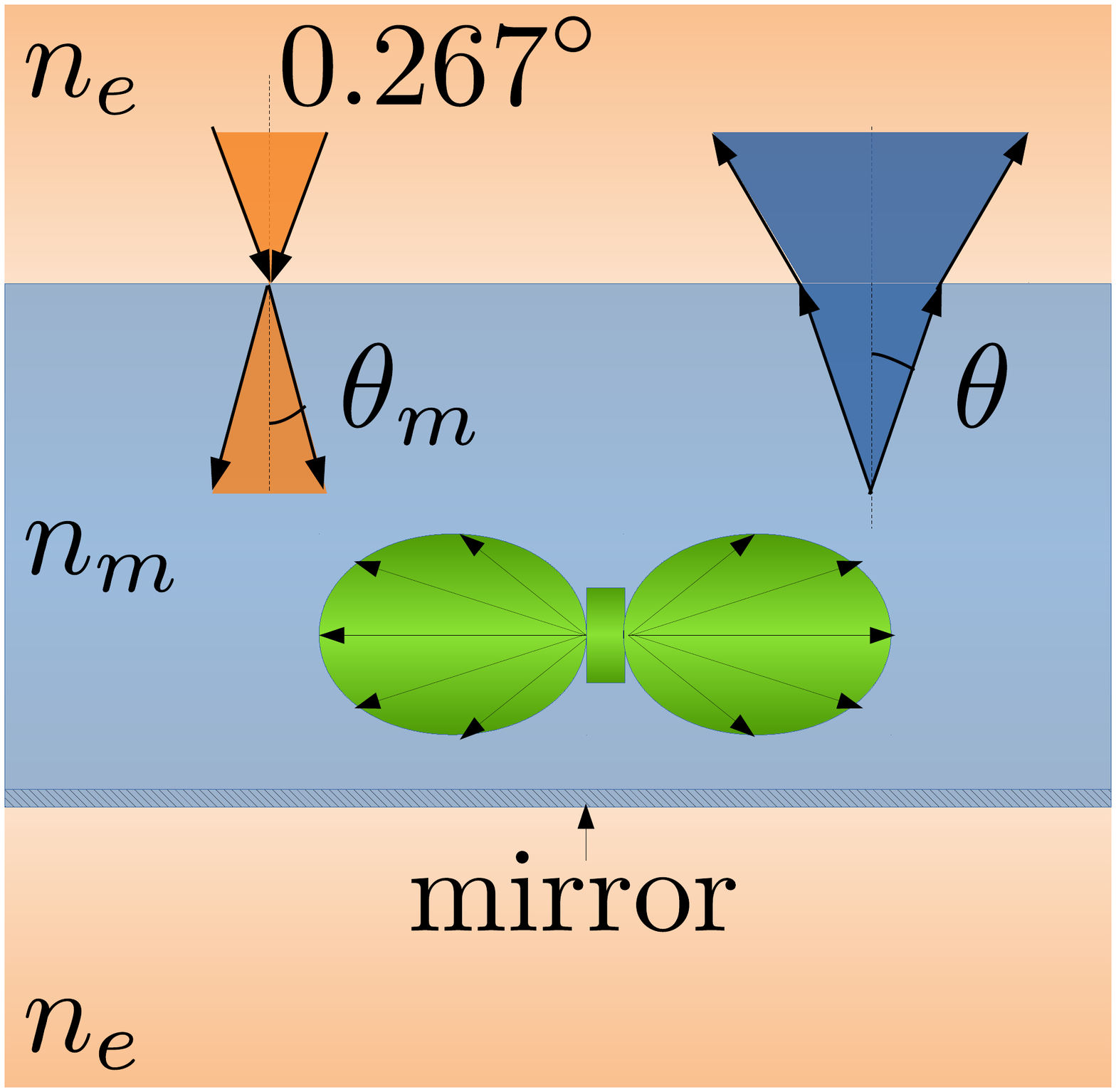}
     } & \subfloat[\label{kirchhoff plot dip noneq}]{%
       \includegraphics[width=0.29\textwidth]{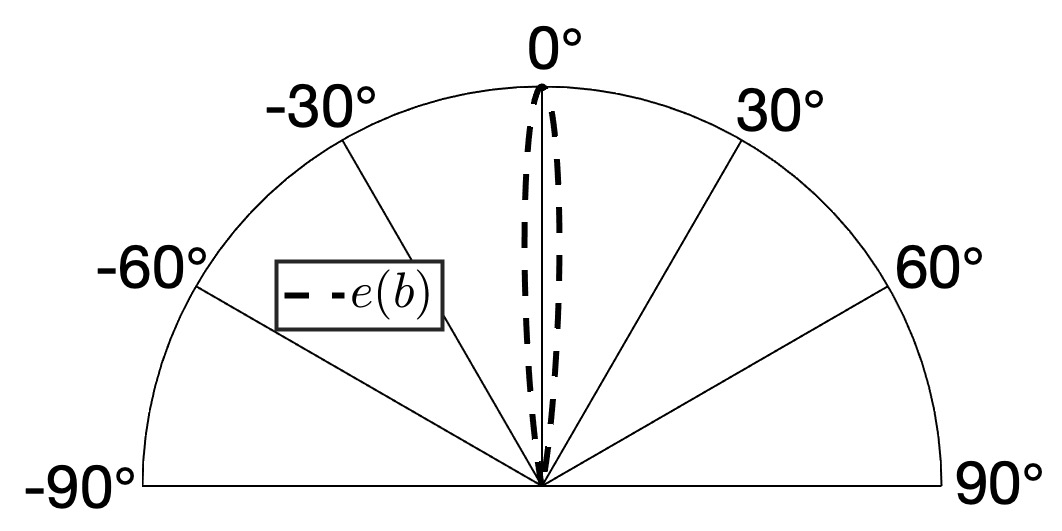}
     } \\
\end{tabular}
\caption{Kirchhoff's Law Analysis. The nonequilibrium case.  (a) and (c) illustrate the two types of media considered, one that emits isotropically and another that emits like a dipole.  For both (a) and (c) the orange semicircle and orange structures represent the incident light whereas the dark blue represents the light exiting the medium.  Light enters the medium in the angular range $0\leq\theta\leq\theta_{m}=\sin^{-1}([n_{e}/n_{m}]\sin(0.267^{\circ}))$.  (b) is a polar plot of the angular distribution of light leaving the medium in the steady state for the isotropic emitting medium.  (d) is the same as (b) except for the dipole emitting case. Incident light, not represented in the plots, spans the solar angle $0\leq\theta\leq 0.267^{\circ}$.  One readily observes that the isotropic emitters emit with a Lambertian radiation pattern unlike the dipolar emitters.}
     \label{fig:kirchhoff's law out of equilibrium isotropic and dipole}
\end{figure}

%Similar to the angle-dependent absorption, we define the angle-dependent emission as
%\begin{align}\label{angle-dependent emission}
    %e(\theta_{b}) &= \frac{1}{d^{(s)}_{0}\sin\theta_{b}}\nonumber\\
    %\times&\left(d^{(s)}_{0}\int_{\theta_{b}-\Delta\theta/2}^{\theta_{b}+\Delta\theta/2}dp^{(e)}_{env}(\theta)p^{(t)}_{env}(\theta,0,0)p^{(a)}_{env}\right.\nonumber\\
    %+&\left.\sum_{i=1}^{M}d^{(s)}_{i}\int_{\theta_{b}-\Delta\theta/2}^{\theta_{b}+\Delta\theta/2}dp^{(e)}(\theta)p^{(t)}(\theta,0,j)p^{(a)}_{env}\right).
%\end{align}

Eqs.~(\ref{angle-dependent absorption}) and (\ref{angle-dependent emission}) are plotted for the isotropic fluorophores and the dipole fluorophores in Figs.~\ref{kirchhoff plot iso} and \ref{kirchhoff plot dip}, respectively, for the case of an isotropic incident radiation field (i.e. the equilibrium state).  One sees that the angle-dependent absorption and the angle-dependent emission are coincident which implies that Kirchhoff's law is upheld.  

%\begin{figure}
%    \centering
%    \includegraphics[width = %\columnwidth]{figures/kir_iso_dip_new_nov7_2020_cropped.png}
%    \caption{Angular distribution for isotropic and dipole angle-dependent emission as a function of angle with solar incident radiation.  Incident light, not represented in the figure, spans the solar angle $0\leq\theta_{e}\leq 0.267^{\circ}$.  One readily observes that the isotropic emitters emit with a Lambertian radiation pattern unlike the dipolar emitters.}
%    \label{fig:kirchhoff's law out of equilibrium isotropic and dipole}
%\end{figure}

In contrast, the non-equilibrium case shows that the angular distributions of the incoming and outgoing light does not match, i.e. $a(b)\neq e(b).$  This can be seen in Fig.~\ref{fig:kirchhoff's law out of equilibrium isotropic and dipole} where the incoming light spans $0\leq\theta\leq 0.267^{\circ}$ (not pictured). Light exiting the medium for the isotropic case is shown in Fig.~\ref{kirchhoff plot iso noneq} while the exiting light for the aligned dipolar fluorophores it is shown in Fig.~\ref{kirchhoff plot dip noneq}.  It is commonly  assumed that most bulk materials fluoresce in a Lambertian manner, like equilibrium radiation fields. We can see that this happens  when the system is composed of isotropic fluorophores, but not when it is composed of aligned dipolar ones, at least with the thickness of medium that we consider. Interestingly, the anisotropy of the emission in the dipolar case is driven more by the lack of reabsorption along certain angles, rather than by the excess emission along other angles.

\subsection{Light Trapping in Luminescent Solar Concentrators}\label{results lsc}

Luminescent solar concentrators (LSCs) have been proposed as an alternative to conventional lens-based optical concentrators for improving the performance and lowering the cost of high efficiency photovoltaic (PV) solar converters, see Ref.~\cite{Martin2020} and references therein. An LSC consists of fluorophores such as dye molecules or semiconductor nanocrystals embedded in a dielectric medium. Light is absorbed by the fluorophores and re-emitted. Refractive index contrast between the medium and environment means that some re-emitted light is trapped by total internal reflection into ``waveguide modes''. PV elements are then placed at the edges of the LSC, where light in the waveguide modes is directed.

The performance of an LSC, and the consequent PV conversion efficiency, depends on the ratio of light occupying waveguide modes compared to the incident sunlight. A useful proxy for this value is the fraction of light each fluorophore emits into the waveguide mode, $\eta_{trap}$, which for isotropic emitters is $\sqrt{1-(n_e/n_m)^2}$ or $\sim 74\%$  for a typical organic polymeric waveguide medium ($n_m/n_e$ = 1.5) \cite{Hermann1982}.  Anisotropic fluorophores, such as aligned semiconductor nanorods, have been proposed as fluorophores because they emit preferentially into the waveguide modes, and can thus provide a higher $\eta_{trap}$. However, a full analysis of LSC performance based on either fluoruphore should also account for the efficiency of absorption of incident light, as well as further re-absorption and re-emission events. Monte Carlo simulations of three-dimensional structures have been used to account for these complexities, as well as scattering and non-unity quantum yield \cite{Leow2013,Bronstein2014,Papakonstantinou2015}. Markov models of the sort described in this paper can make this fuller analysis more computationally tractable.

An LSC can be described with the same geometry  analyzed in Sec.~\ref{results kirchhoffs law}, with a mirror on the bottom of the medium. We assume that the fluorophores are suspended in a solid matrix of refractive index $1.49$, while the refractive index of the environment is 1. We consider both isotropic fluorophores, as in  Fig.~\ref{kirchhoff ill iso}, and dipolar as in Fig.~\ref{kirchhoff ill dip}. We assume diffuse incident light with a Lambertian profile. We do not track the wavelength of light specifically, but to account for a Stokes shift, we use an absorption coefficient for incident solar radiation of $5\times 10^8 \textrm{cm}^{-1}\textrm{M}^{-1}$, while we use $5\times 10^5 \textrm{cm}^{-1}\textrm{M}^{-1}$ ( $\sim$ 1000x lower cross section) for light emitted from the fluorophores, in agreement with common values for semiconductor nanocrystals \cite{Martin2020}.

We can use the previous analysis to find the steady state population of excited fluorophores in the medium, $\sum_{i=1}^M d_i/d_0$. The trapping ratio of re-emitted light is $\eta_{trap}=\int_{\theta_c}^{\pi/2}dp^{(e)}(\theta)$ (which is $\sqrt{1-(n_e/n_m)^2}$ for isotropic emitters). With quantum yield $\eta_Q$, the net occupation of the waveguide modes, which we will call the ``occupation factor'', is given by the product of these three factors, \begin{equation}\label{trapping efficiency}
    \mathcal{O}=\eta_Q\eta_{trap} \sum_{i=1}^M\frac{d_i}{d_0}.\end{equation}

Because our model assumes an infinite horizontal extent of the medium, it does not directly track how much light reaches the photovoltaics at the edges. Instead it identifies how much light occupies the waveguide modes at any point in the medium, just as the analysis based on $\eta_{trap}$ alone does. To fully track the three-dimensional motion of the photons from initial incidence until PV conversion, the Markov model would require states representing differential volume elements, rather than thin layers, as well as transition probabilities between them, but the process of finding the stationary distribution would be the same.

\begin{figure}[ht]
    \centering
    \includegraphics[width=0.45\textwidth]{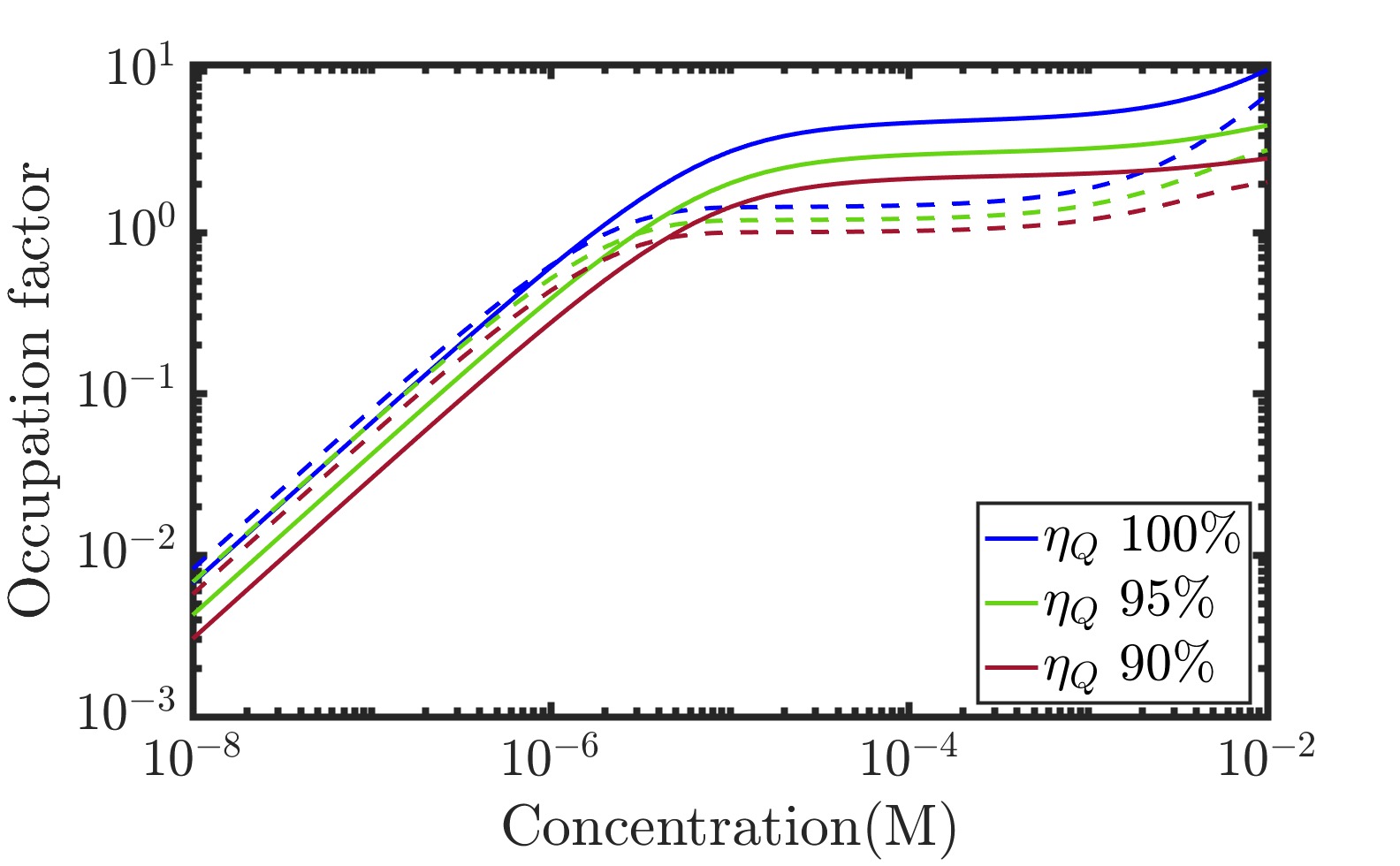}
    \caption{Occupation of  waveguide modes compared to incident photons, as a function of concentration of fluorophores in the LSC. Solid traces correspond to dipolar fluorophores while dashed traces correspond to isotropic fluorophores.}
    \label{fig:LSC_figure}
\end{figure}

Figure~\ref{fig:LSC_figure} plots the occupation factor with dipolar and isotropic fluorophores and variable quantum yield, as a function of the concentration of nanocrystals in the polymer matrix of the LSC, spanning a range commonly studied experimentally. Values greater than 1 can be achieved because a single photon may be absorbed and re-emitted several times. Note that dipolar fluorophores achieve higher occupation factors when concentration is high, due in part to higher $\eta_{trap}$, while they have lower occupation factor at low concentrations due to absorbing less incident light than isotropic fluorophores.

\section{Conclusion}

We have demonstrated the use of a Markov model to study the steady state behavior of several optical systems. The method works by finding an eigenvector of the matrix of transition probabilities for photons moving through the system. We have shown that this method yields the expected results in systems that allow for an analytical solution (Beer-Lambert's law or Kirchhoff's law). This model also gives a description of steady states for which analytical solutions are not available. We propose that this method could be particularly useful for describing complex mesoscale interactions between nanophotonic elements, such as in an LSC.

%We have modeled fluorescence, scattering, and absorption in several optical systems, some of which were composed to nanoscale emitters.  For the optical system considered, the path of a photon after multiple absorption and emission/scattering events can be thought of a random process, like a random walk.  We use the Markov chain to predict where they photon may end up in the long run, i.e. its steady state distribution.  We have demonstrated the our model replicate physical laws such as Beer-Lambert's law and Kirchhoff's law of reciprocity.  We applied the Markov chain model to two luminescent solar concentrators and simulated three loss pathways associated with these devices.

The particular model we developed assumes translational symmetry in two dimensions, and only tracks the movement of photons in the third dimension. This model also assumes that the angle of emission of a photon is independent of the angle of absorption, which makes less accurate predictions when applied to analysis of scattering. However, more detailed Markov models can track these additional variables to give more precise results. Transition probabilities can also be modified to take into account additional features of radiation, such as near-field interactions, wavelength dependence, and deviation from perfect isotropic or sine-squared emission from the fluorophores.

%\section{References}
\bibliographystyle{naturemag}
\bibliography{references}

\end{document}